\documentclass[12pt]{iopart} 

\usepackage{subfig}

\expandafter\let\csname equation*\endcsname=\relax
\expandafter\let\csname endequation*\endcsname=\relax
\usepackage{amsmath}
\usepackage{graphicx}
\usepackage{iopams,cite}  
\begin{document}

\title[Heine-Stieltjes and Van Vleck polynomials associated with integrable BCS models]{Generalised Heine-Stieltjes and Van Vleck polynomials associated with degenerate, integrable BCS models}

\author{Ian Marquette and Jon Links}

\address{School of Mathematics and Physics, The University of Queensland, Brisbane, QLD 4072, Australia}
%\ead{i.marquette@uq.edu.au, jrl@maths.uq.edu.au}

\eads{\mailto{i.marquette@uq.edu.au}, \mailto{jrl@maths.uq.edu.au}}

\begin{abstract}
We study the Bethe Ansatz/Ordinary Differential Equation (BA/ODE) correspondence for Bethe Ansatz equations that belong to a certain class of coupled, nonlinear, algebraic equations. Through this approach we numerically obtain the generalised Heine-Stieltjes and Van Vleck polynomials in the degenerate, two-level limit for four cases of exactly solvable Bardeen-Cooper-Schrieffer (BCS) pairing models. These are the $s$-wave pairing model, the $p+ip$-wave pairing model, the $p+ip$ pairing model coupled to a bosonic molecular pair degree of freedom, and a newly introduced extended $d+id$-wave pairing model with additional interactions. The zeros of the generalised Heine-Stieltjes polynomials provide solutions of the corresponding Bethe Ansatz equations. 
We compare the roots of the ground states with curves obtained from the solution of a singular integral equation approximation, which allows for a characterisation of ground-state phases in these systems. Our techniques also permit for the computation of the roots  of the excited states. These results illustrate how the BA/ODE correspondence can be used to provide new numerical methods to study a variety of integrable systems. 
\end{abstract}

%\ams{81R05}

\maketitle

\section{Introduction}

The correspondence between zeros of polynomials and models of physical systems has a very long history which goes back to early works of  Stieltjes \cite{Sti}, Heine \cite{Hei}, B\^ocher \cite{Boc}, Van Vleck \cite{Van} and Polya \cite{Pol} at the end of the 19th century in relation to electrostatic models. A concise summary can be found in \cite{Mar}. The subject of zeros of polynomials \cite{Sze}, and specifically the sum rules of their zeros, was systematically studied  for Hermite, Laguerre, Tchebycheff, Jacobi and Lam\'e polynomials \cite{Cas1,Cas2}. The study of the generalised Lam\'e equation (a second order Fuchsian equation), its Heine-Stieltjes and Van Vleck polynomials, and their zeros, is still a very active area of research \cite{Bou1,Bou2,Berg,Sha1,Sha2,Sha3,Hol,z12}. 

The relation between ordinary differential equations and integrable systems has been observed by several authors \cite{Eno,Uly,Dor1,blz,Dor2} 
%In some of these papers, the connection involved differential equations of the Schr\"odinger form with quasi-exactly solvable %potentials \cite{Eno,Uly,Dor1,blz,Dor2}. 
and in particular connection to a generalised Stieltjes problem  has been discussed  \cite{Sri,Bat}. Richardson's Bethe Ansatz solution \cite{Ric1,Ric2} for the $s$-wave pairing Bardeen-Cooper-Schrieffer (BCS) Hamiltonian has attracted considerable attention \cite{Gau,Cam,vdr,zlmg02,vdp02,Roma,Ami2,o03,yba05,Dom,fcc09,fcc10,Far,cc11,p12}. The correspondence between the Richardson's equations and the confluent Heun differential equation was recognised by Gaudin \cite{Gau}. Numerical methods to solve Richardson's equations for finite number of particles  were implemented later \cite{Roma}. Here there may be convergence problems  concerning critical points where the solution set of roots contains degeneracies \cite{Dom}. The numerical methods of \cite{Roma,Dom} rely on a direct approach, or use an appropriate change of variables for the critical points, to solve the Bethe Ansatz equations.

With regard to the difficulty of directly solving the Bethe Ansatz equations, very recently new numerical approaches based on the Bethe Ansatz/Ordinary Differential Equation (BA/ODE) correspondence were proposed \cite{Far,Ara,Pan1,Pan2}. They are based on linear 
second-order differential equations \cite{Pan1,Pan2}, or their corresponding Riccati equations \cite{Far,Ara} which are first-order nonlinear differential equations. The polynomials obtained by these methods correspond to extended Heine-Stieltjes polynomials \cite{Sri}. Many advantages of these approaches were discussed in these works and it appears that such methods could be beneficial for studying integrable systems at their critical points. However, the applications of these new methods were limited to a certain class of Bethe Ansatz equations that contained the Richardson solution and the Lipkin-Meshkov-Glick (LMG) model. 

In recent years, many papers have appeared devoted to finding new examples of BCS systems solvable by the Bethe Ansatz \cite{Dun1,Anf,dorvh06, Iba,s09,s09a,Dun2,Romb,Duk1,Dun3,hlzw11,Wu,Bir}. In contrast to the Richardson solution for $s$-wave pairing, several of these newer systems exhibit quantum phase transitions which can be identified by a change in the character of the Bethe roots as a coupling constant is varied. Such behaviour has also recently been observed in a different context of bosonic models \cite{rfmr12}. The purpose of the present paper is to provide a first step towards extending the application of numerical methods based on the BA/ODE correspondence to these more recent BCS systems. In particular we will conduct an analysis for the degenerate two-level limit.

The BCS Hamiltonians we consider below are generally expressed in terms of fermions. Given  the blocking effect \cite{vdr} of the pairing interaction, for simplicity we hereafter  only consider for all models the subspace of unblocked states. This effectively reduces the dimension of the Hilbert space of states for the model. In this case it is more convenient to write the Hamiltonians only in term of hard-core boson operators.
The hard-core boson operators $b_{j}^{\dagger}$ and $b_{k}$ (Cooper pair creation and annihilation operators) satisfy the following commutation relations
\begin{equation}
(b_{j}^{\dagger})^{2}=0,\quad [b_{j},b_{k}^{\dagger}]=\delta_{jk}(1-2N_j),\quad [b_{j},b_{k}]=[b_{j}^{\dagger},b_{k}^{\dagger}]=0 \label{eq12}
\end{equation}
with $N_j=b_{j}^{\dagger}b_{j}$.
Furthermore by considering a degenerate, two-level limit of these models we further effectively reduce the dimensionality of the Hilbert space by only considering certain ``unblocked symmetric states'' which will be defined later. One motivation for us to do this is that we want to illustrate  that the numerical methods employed here provide solutions of the Bethe Ansatz equations for excited states as well as the ground state. In order to display this data in a transparent manner, it is an advantage to restrict our analysis to a subspace of the full Hilbert space. If the ground state on the unrestricted Hilbert space is non-degenerate, symmetry arguments lead to the conclusion that the ground state is an unblocked symmetric state. Finally, we mention that the method employed here (and in \cite{Pan1,Pan2}) is distinct from other approaches \cite{Roma,Dom,Far,Ara,Dun2,Romb} in that we do not track solutions of the Bethe Ansatz equations from the zero coupling limit. In this respect issues surrounding critical points are avoided.   

In Section 2 we will introduce a class of nonlinear algebraic equations and obtain the corresponding ordinary differential equations  involving only polynomials. We classify the different cases and discuss the application to two-level models. In Section 3, we discuss an algorithm to numerically generate the corresponding generalised Heine-Stieltjes and Van Vleck polynomials. We apply this approach to four systems: the Richardson $s$-wave model \cite{Ric1}, the $p+ip$-wave pairing model \cite{Iba,Dun2,Romb}, the $p+ip$-wave pairing model coupled to a bosonic molecular pair degree of freedom \cite{Dun3}, and a new solution for a $d+id$-wave pairing model with additional interactions \cite{ml12}. We will compare the ground-state roots obtained by this numerical method with the theoretical curve for an arc obtained in the continuum limit by a singular integral equation. In two cases we observe a phenomenum whereby all the ground-state roots collapse at the origin at a particular value of the coupling parameter. The method also allows us to obtain the roots for the excited states, which is important for studies of the dynamics analogous to those in \cite{fcc09,fcc10}. 

\section{The generalised Heine-Stieltjes correspondence}

The systems studied previously using the polynomial approach \cite{Far,Ara,Pan1,Pan2} belong to the class of coupled nonlinear algebraic (Bethe Ansatz)  equations of the form

\begin{equation}
\sum_{i=1}^{L}\frac{\rho_{i}}{y_{l}-\varepsilon_{i}}-2\sum_{j\neq l}^{M} \frac{1}{y_{l}-y_{j}}+C+D y_{l}=0 , \qquad l=1,...,M \label{eq1}
\end{equation}
where $C,D,\rho_{i}$ and $\varepsilon_{i}$ are real parameters. The parameters $\varepsilon_{i}$ can be interpreted in the context of BCS systems as single particle energy levels, $L$ is the number of levels and $M$ is the total number of Cooper pairs. The various systems studied in  \cite{Iba,Dun2,Romb,Duk1,Dun3,Bir} have Bethe Ansatz equations that belong to a different class than the one given by Eq. (\ref{eq1}), taking the following form

\begin{equation}
\sum_{i=1}^{L}\frac{\rho_{i}}{y_{l}-\varepsilon_{i}}-2\sum_{j\neq l}^{M} \frac{1}{y_{l}-y_{j}}+\frac{A}{y_{l}^{2}}+\frac{B}{y_{l}}+C=0,
\qquad l=1,...,M. \label{eq2}
\end{equation}
In the equation above, the real parameters $A$, $B$ and $C$ may depend on $L$ and $M$ or other constants depending of the nature of the physical problem considered. 

We start by constructing the following polynomials

\begin{equation}
Q(z)=\prod_{j=1}^{M}(z-y_{j}),\quad P(z)=\prod_{i=1}^{L}(z-\varepsilon_{i}). \label{eq3}
\end{equation}
It is well  known that such polynomials satisfy the following relations
\begin{equation}
\frac{Q''(y_{l})}{Q'(y_{l})}=2\sum_{j\neq l}^M\frac{1}{y_{l}-y_{j}} ,\quad \frac{P'(y_{l})}{P(y_{l})}=\sum_{i=1}^L\frac{1}{y_{l}-\varepsilon_{i}} .\label{eq4}
\end{equation}
We introduce a polynomial $W(z)$ which is of order $L-1$ such that for the set of real parameters $\rho_{i}$ and $\varepsilon_{i}$ it satisfies
\begin{align*}
\frac{W(z)}{P(z)}=\sum_{i=1}^{L}\frac{\rho_{i}}{z-\varepsilon_{i}}, %\label{eq5}
\end{align*}
and in the case $\rho_{i}=1$ $\forall i$, as can be seen from Eq.(\ref{eq4}), $W(z)=P'(z)$. These relations can be used to construct from Eq. (\ref{eq2}) the following differential equation
\begin{align*}
\frac{A}{y_{l}^{2}}+\frac{B}{y_{l}}+C+\frac{W(y_{l})}{P(y_{l})}-\frac{Q''(y_{l})}{Q'(y_{l})}=0. %\label{eq6}
\end{align*}
This equation can be written in the form $A_{2}(y_l)Q''(y_l)+A_{1}(y_l)Q'(y_l)=0$.  
Because, the polynomial $Q(z)$ vanishes at the solutions $y_l$ of the Bethe Ansatz equations we can thus form the following second order differential equation
\begin{equation}
A_{2}(z)Q''(z)+A_{1}(z)Q'(z)=A_{0}(z)Q(z), \label{eq7}
\end{equation}
where the order $|A_{j}|$ of the polynomial  $A_{j}(z)$ depends on when the parameters $A,B,C$ vanish. Seven cases can occur, as given in  Table 1.

\begin{table}
\caption{Cases of generalised Heine-Stieltjes and Van Vleck polynomials for the differential equation given by Eq. (\ref{eq7})}
\footnotesize\rm
\begin{tabular*}{\textwidth}{@{}l*{15}{@{\extracolsep{0pt plus12pt}}l}}
\br
 Case & A   &   B   &   C  &  $A_{2}$ &  $A_{1}$ & $|A_{2}|$ & $|A_{1}|$ & $|A_{0}|$  \\
\br
  1 & $A \neq 0$ & $B \neq 0$ & $C \neq 0$ & $z^{2}P$   & $(-A-Bz-Cz^{2})P-z^{2}W$ & $L+2$ & $L+2$  & $L+1$  \\
  2 & $A \neq 0$ & $B \neq 0$ & $C = 0$ &   $z^{2}P$ & $(-A-Bz)P-z^{2}W$ &  $L+2$ & $L+1$ & $L$ \\
  3 & $A \neq 0$ & $B=0$  & $C \neq 0$ & $z^{2}P$  & $(-A-Cz^{2})P-z^{2}W$ &  $L+2$ & $L+2$ & $L+1$ \\
  4 & $A \neq 0$ & $B=0$  & $C = 0$ & $z^{2}P$  & $-AP-z^{2}W$ &  $L+2$ & $L+1$ & $L$ \\
  5 & $A =0$ & $B \neq 0$ & $C \neq 0$ & $zP$ & $(-B-Cz)P-zW$ & $L+1$ & $L+1$ & $L$ \\
  6 & $A = 0$ & $B \neq 0$ & $C = 0$ & $zP$ & $-BP-zW$ & $L+1$ & $L$ & $L-1$ \\
  7 & $A = 0$ & $B = 0$ & $C \neq 0$ & $P$ & $-CP-W$ & $L$ & $L$ & $L-1$ \\
\mr
\end{tabular*}
\end{table}

For a given $L$, the polynomial $A_{0}(z)$ can be seen as a generalised Van Vleck polynomial and the polynomial $Q(z)$ as a generalised Heine-Stieltjes polynomial.  For our numerical investigations we will study degenerate two-level models, where each of the distinct levels $\varepsilon_1$ and $\varepsilon_2$ have have degeneracy $L/2$. However the method can be applied to general $L$-level models at the cost of much more involved numerical calculations. For our study the problem becomes mathematically equivalent to taking $\rho_{i}={L}/{2}$ $\forall i$ for the first term of Eq. (\ref{eq2}). The Bethe Ansatz equations then take the following form 

\begin{equation}
\frac{A}{y_{l}^{2}}+\frac{B}{y_{l}}+C-\frac{L}{2}\left(\frac{1}{\varepsilon_{1}-y_{l}}+\frac{1}{\varepsilon_{2}-y_{l}}\right) -2\sum_{j\neq l}^{M} \frac{1}{y_{l}-y_{j}}=0 .\label{eq8}
\end{equation}

\section{Numerical method based on BA/ODE correspondence and examples}

We will use Eq. (\ref{eq7}) to construct the polynomials $Q(z)$ and $A_{0}(z)$ by inserting expansions of these polynomials and solving the corresponding system of equations. From the coefficients of the polynomial $Q(z)$ we obtain the roots by standard techniques, and they correspond to the solutions of the corresponding Bethe Ansatz equations. This approach is similar to the one taken in \cite{Pan1,Pan2}. 

We start by taking the following expansions (where $M$ and $L$ are fixed)
\begin{equation}
Q(z)=\sum_{j=0}^{M}\alpha_{j}z^{j},\quad A_{0}(z)=\sum_{j=0}^{K}\beta_{j}z^{j}, \label{eq9}
\end{equation}
with $\alpha_{M}=1$ and the remaining $\alpha_{j}$ and $\beta_{j}$ are coefficients to be determined numerically. Inserting the expansions given by Eq. (\ref{eq9}) into Eq. (\ref{eq7}) yields two matrix equations that we need to solve. The coefficients of $z^{i}$ for $i=0,...,M$ generate a $(M+1)\times(M+1)$ matrix $F$ that satisfies the equation $F{\mathbf v}=\beta_{0}{\mathbf v}$ with  ${\mathbf v}^T=(\alpha_{0},...,\alpha_{M})$. The coefficients of $z^{i}$  for $i=M+1,...,M+K$ will generate a $K\times(M+1)$ upper triangular matrix $P$ that satisfies  
$P{\mathbf v}={\mathbf 0}$. The matrix entries are all linear in the  coefficients of the generalised Van Vleck polynomials $\{\beta_{1},...,\beta_{K}\}$. We obtain a set of coefficients $\alpha_{k}$ (with $k \in [0,M]$) for each solution set of the Bethe Ansatz equations and a corresponding set of coefficients $\beta_{j}$ (with $j \in [1,K]$). Some of the $\beta_{j}$  may be independent of the $\alpha_{k}$  and thus do not depend on the set of Bethe roots. We have adapted a Mathematica code discussed in \cite{Pan1} to implement the calculations.  

An effect of restricting to the unbloacked symmetric states of the degenerate, two-level models which are governed by  Eq. (\ref{eq8}) is that the order of the polynomials $A_{0}$, $A_{1}$ and $A_{2}$ is fixed as we change the value of $L$ and $M$. The polynomials are respectively of order 3, 4 and 4 at most, and thus the parameter $K$ is at most 3: 
\begin{align*}
A_{1}(z)=\sum_{j=0}^{4}\xi_{j}z^{j},\, A_{2}(z)=\sum_{j=0}^{4}\zeta_{j}z^{j}. %\label{eq10}
\end{align*}
When $A_{1}$ reduce to a polynomial of order 3, the equation fall into the class of differential equations recently studied in \cite{z12}.
Once the $\alpha_{i}$ and $\beta_{j}$ are obtained, each solution corresponding to solution set of Bethe roots, the next step is to compute the roots from the generalised Heine-Stieltjes polynomials. This can be done in principle using one of the many well-known methods implemented in available softwares, and for an arbitrary order. But in practice this is a more complicated issue. A difficulty that appears is that these generalised Heine-Stieltjes polynomials will be polynomials of order $M$ and care is needed as it is known that numerical methods to find roots of polynomials can have instabilities \cite{Wil,Cor,Far4}.  This problem can even affect the structure of the roots (i.e. real vs. complex conjugate pairs). 
%
%This problem can be adressed in many ways. One of them is to use a sufficient numerical precision, or we can also use a basis other than %the monomial basis for the expansion of the polynomials. Two types of polynomials have been considered the barycentric ( refered also in %the literature as Lagrange or Wilkinson) polynomials and the Bernstein polynomials \cite{Bern} introduced 100 years ago. From these %works \cite{Wil,Far1,Cor,Far2,Rab,Far3,Far4,Bern} it was pointed out that conversion between basis can be sometimes unstable and that if %the Bernstein polynomials are the optimal basis, the Lagrange polynomial can be sometime more stable and thus more appropriate. However, %the non negativity of Bernstein polynomial can be an advantage in certain circumstances. This issue was observed in \cite{Pan1} and it %was proposed that specific codes could be developped in this context instead of using methods implemented in Mathematica. This issue was %adressed in Ref.33 using barycentric polynomials to calculate the roots instead of the monomial expansions. However, it was observed %that the method remain heavily sensible to finite numerical precision if we do not provide a grid sufficiently close to the actual %roots. 
%
We will adopt in this paper an approach based on monomial expansion but we will take care of using a sufficient working precision. We will show that the method appears to be reliable and robust. We will present explicit examples of polynomials in the Appendix A, illustrating that the coefficients can be of very different orders of magnitude.

%This is also necessary to classify the states. In \cite{Pan1} and \cite{Pan2}, the cases studied had energy given as the sum of the %roots the ground state. The ground state can be identified simply by considering $Min[-a_{M-1}]$. In the cases we will discuss one of %them have energy that depend also of the double sum of the roots. This is thus necessary to classify the term using the explicit formula %for the energy, the ground is the state with the lowest energy.

\subsection{\textbf{Example 1: Richardson $s$-wave pairing model }}

We first consider the exactly solvable $s$-wave pairing BCS model \cite{Ric1,Ric2,Gau,Cam,vdr,Roma,Ami2,zlmg02,Dom,o03,vdp02,yba05} given by the following Hamiltonian expressed in terms of the hard-core boson operators
\begin{align*}
H=\sum_{j=1}^{L}\varepsilon_{j}N_{j}-G\sum_{j,k}^{L}b_{j}^{\dagger}b_{k}. %\label{eq11}
\end{align*}
Omitting blocked states, this Hamiltonian acts on a Hilbert space of dimension $2^L$. 
The energy is given by
\begin{equation}
E=\sum_{j=1}^{M}y_{j}, \label{eq15}
\end{equation}
where the roots $y_{j}$ satisfy Eq. (\ref{eq2}) with $C={G}^{-1}$ and $A=B=0$. This corresponds to  case 7 of Table 1.

Imposing $L$ is even the degenerate two-level limit is obtained by setting 
\begin{align*}
\varepsilon_j=\begin{cases}
\varepsilon_1 \quad j\,\,{\rm odd},\\
\varepsilon_2 \quad j\,\,{\rm even}
\end{cases}
\end{align*} 
in which case the Hamiltonian may be expressed as 
\begin{align*}
H=L+2\varepsilon_1 {\mathcal S}^z_1+2\varepsilon_2 {\mathcal S}^2_2
-G({\mathcal S}_1^++{\mathcal S}_2^+)({\mathcal S}^-_1 +{\mathcal S}^-_2)
\end{align*}
where 
\begin{align}
{\mathcal S}^z_1&= \frac{1}{2}\sum_{j\,\,{\rm odd}}(2N_j-I), &
{\mathcal S}^z_2&= \frac{1}{2}\sum_{j\,\,{\rm even}}(2N_j-I), \label{su21}\\
{\mathcal S}^-_1&= \sum_{j\,\,{\rm odd}}b_j, &
{\mathcal S}^-_2&= \sum_{j\,\,{\rm even}}b_j, \label{su22} \\
{\mathcal S}^+_1&= \sum_{j\,\,{\rm odd}}b^\dagger_j, &
{\mathcal S}^+_2&= \sum_{j\,\,{\rm even}}b^\dagger_j \label{su23}
\end{align}
provide two representations of the $su(2)$ algebra.
Now the energies given by (\ref{eq15}),
%
%\begin{equation}
%E=\sum_{j=1}^{M}y_{j}, \label{eq15}
%\end{equation}
%
where the roots $y_{j}$ satisfy Eq. (\ref{eq8}) with $C={G}^{-1}$ and $D=0$, only provides a subset of the spectrum for the Hilbert space of dimension $2^L$. These correspond to states which are invariant under the mutual interchange of the even subscripts, and the mutual interchange of the odd subscripts, which label the components of the tensor products. We term these {\it unblocked symmetric states}. This subspace has dimension $(L/2+1)^2$, and will be the focus of our study in this section. In representation-theoretic terms, the hard-core operators satisfying Eq. (\ref{eq12}) can be mapped to $L$ irreducible spin-1/2 $su(2)$ representations, whereas  (\ref{su21},\ref{su22},\ref{su23}) are two reducible $su(2)$ representations. The Eq. (\ref{eq8}) with $C={G}^{-1}$ and $D=0$ is only associated with one irreducible component of these reducible representations, that with spin-$L/4$ for each $su(2)$ copy. Similar considerations will apply for all subsequent models that we study.    
%
%
%
%
%The conserved quantities (i.e. the quantum mechanical operators that commute with the Hamiltonian ($[H,\tau_{j}]=0$)) are thus given by
%
%\begin{equation}
%\tau_{j}=-\frac{1}{2G}(N_{j}-I)+\sum_{k\neq j}^{L}\frac{\theta_{jk}}{k_{i}-k_{k}}, \label{eq13}
%\end{equation}
%
%with $\theta$ given by the following product
%\begin{equation}
%\theta=b^{\dagger}\otimes b+b\otimes b^{\dagger}+\frac{1}{2} (N_{j}-I)\otimes (N_{j}-I). \label{eq14}
%\end{equation}
%
%They were obtained by various methods and in particular using the quantum inverse scattering methods \cite{Lin}. 
%

Setting $\gamma=\varepsilon_{1}+\varepsilon_{2}$ and $\eta= \varepsilon_{1}\varepsilon_{2}$, in the degenerate two-level limit the differential equation takes the form 
\begin{align*}
(z^{2}-\gamma z+\eta)Q''+\left(-\frac{1}{G}z^{2}+\left(\frac{\gamma}{G}-L\right)z+\frac{1}{2}\gamma L-\frac{\eta}{G}\right)Q'-(\beta_{1}z+\beta_{0})Q=0. %\label{eq16}
\end{align*}
From the first and second terms in the expansion of the polynomial $Q$ (using Eqs. (\ref{eq3},\ref{eq15})) 
\begin{align*}
Q(z)=z^{M}-Ez^{M-1}+...+(-1)^M\prod_{j=1}^{M}y_{j}, %\label{eq17}
\end{align*}
we have that
\begin{align*}
\beta_{1}= -\frac{M}{G}  ,\quad \beta_{0}=-\left(\frac{\gamma}{G}M+\frac{E}{G}+M+LM-M^{2}\right). %\label{eq18}
\end{align*}
Because the parameter $\beta_{0}$ depends on the energy, we obtain a Van Vleck polynomial corresponding to the Heine-Stieltjes polynomial of each eigenstate of the Hamiltonian.  

We take the case where $\varepsilon_{2}=1$ and $\varepsilon_{1}=-1$, following \cite{Gau,Roma}. We choose $L=100$, a system at half-filling $M=50$, and introduce the scaled coupling constant $g=GL$ which will be used throughout. From the numerical method based on the BA/ODE correspondence we reproduce the ground state results of \cite{Gau,Roma}. We see that the results obtained by this method also agree with the theoretical distribution curve in the large-$L$ limit which is calculated in Appendix B. The case $g=1$ is identified as a critical point at which the character of the arc changes from being a closed curve to an open curve. The method we use also computes the roots of the excited states.  Fig. 2 shows all sets of the Bethe roots for given values of $g$. The number of states in the $M=50$ sector is 51, so the number of data points in each panel is $50\times 51=2550$.

\begin{figure}
\centering
\subfloat[]{\includegraphics[width=5.2cm]{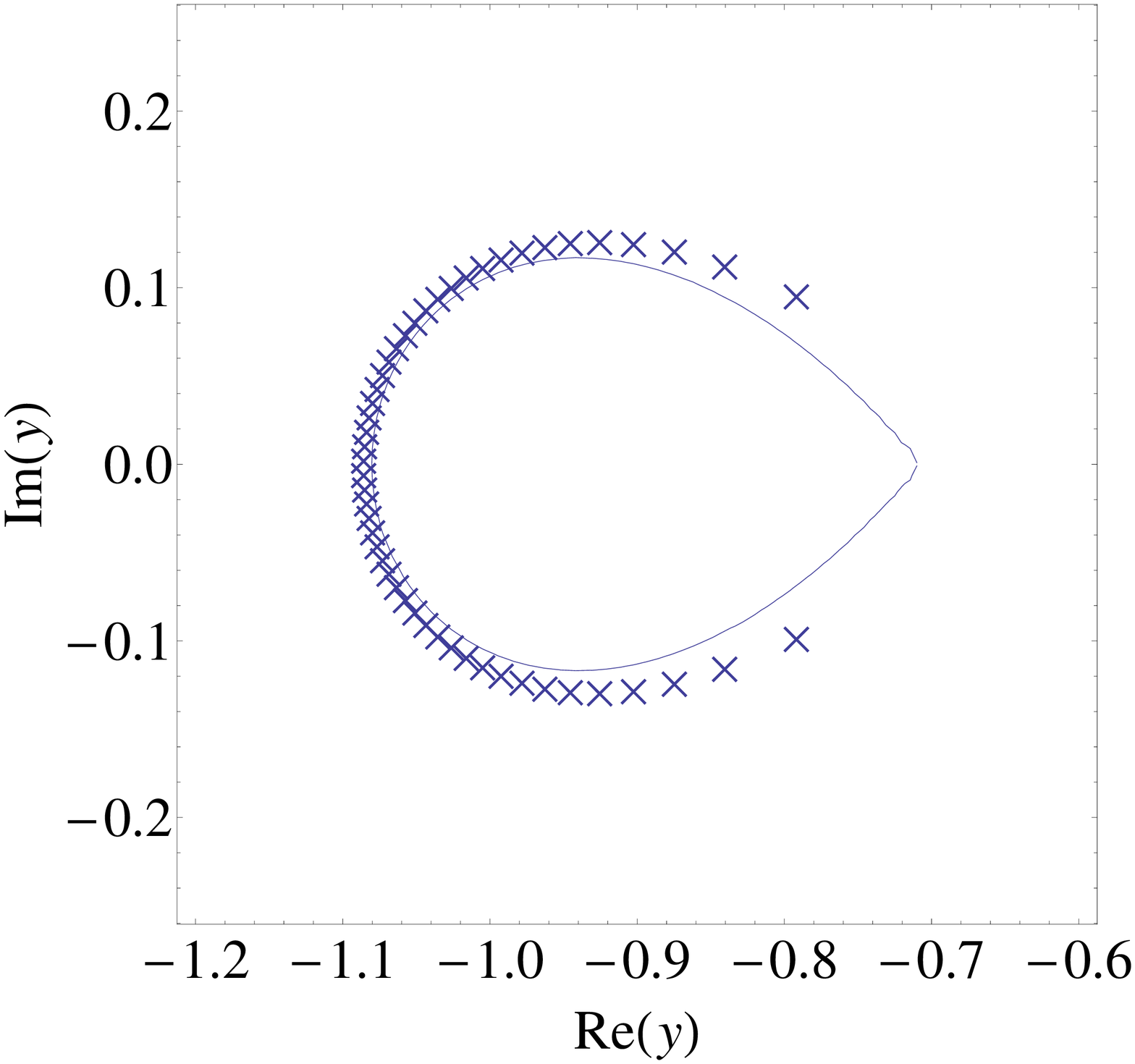}}
\subfloat[]{\includegraphics[width=5cm]{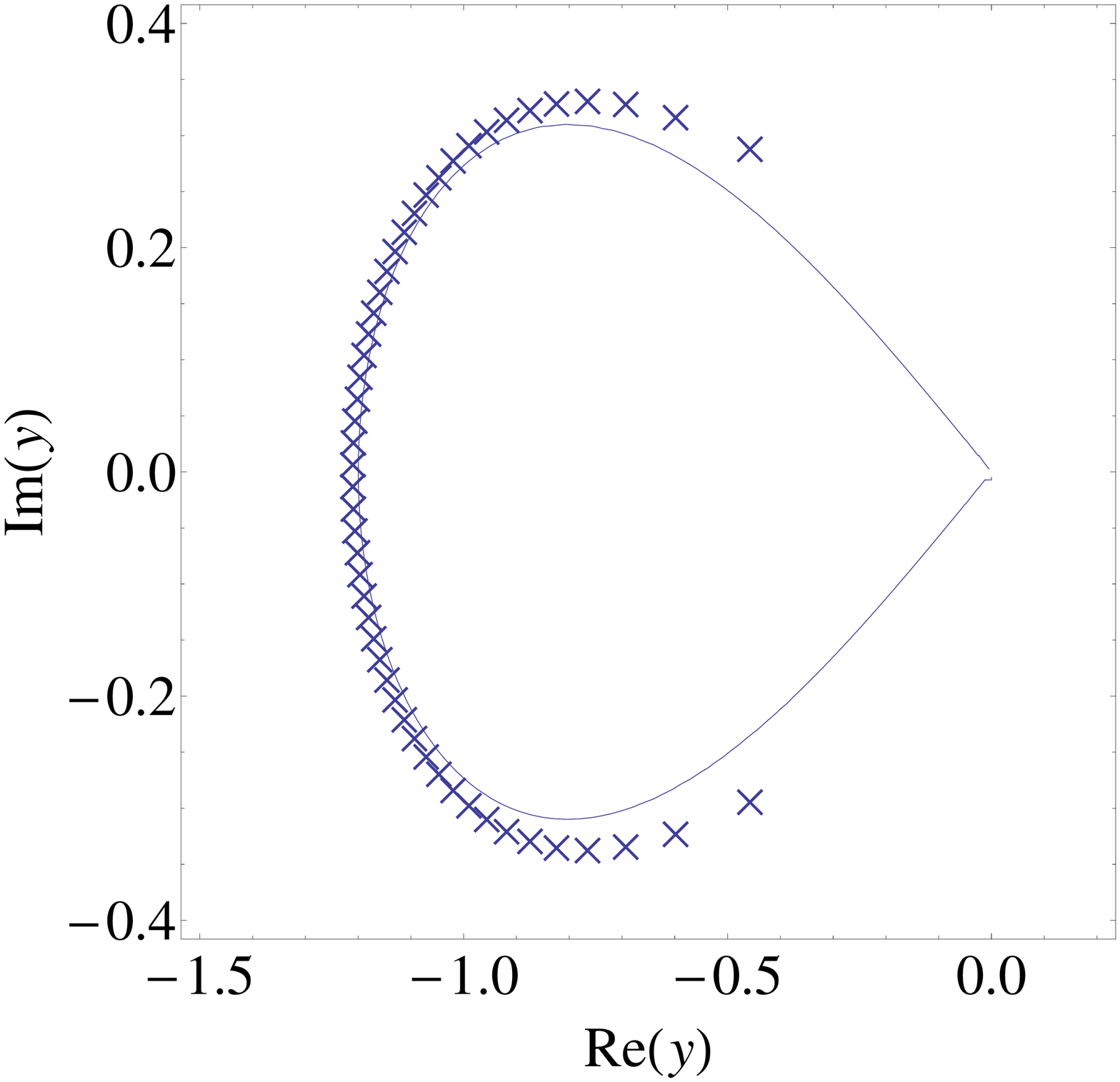}}
\subfloat[]{\includegraphics[width=5cm]{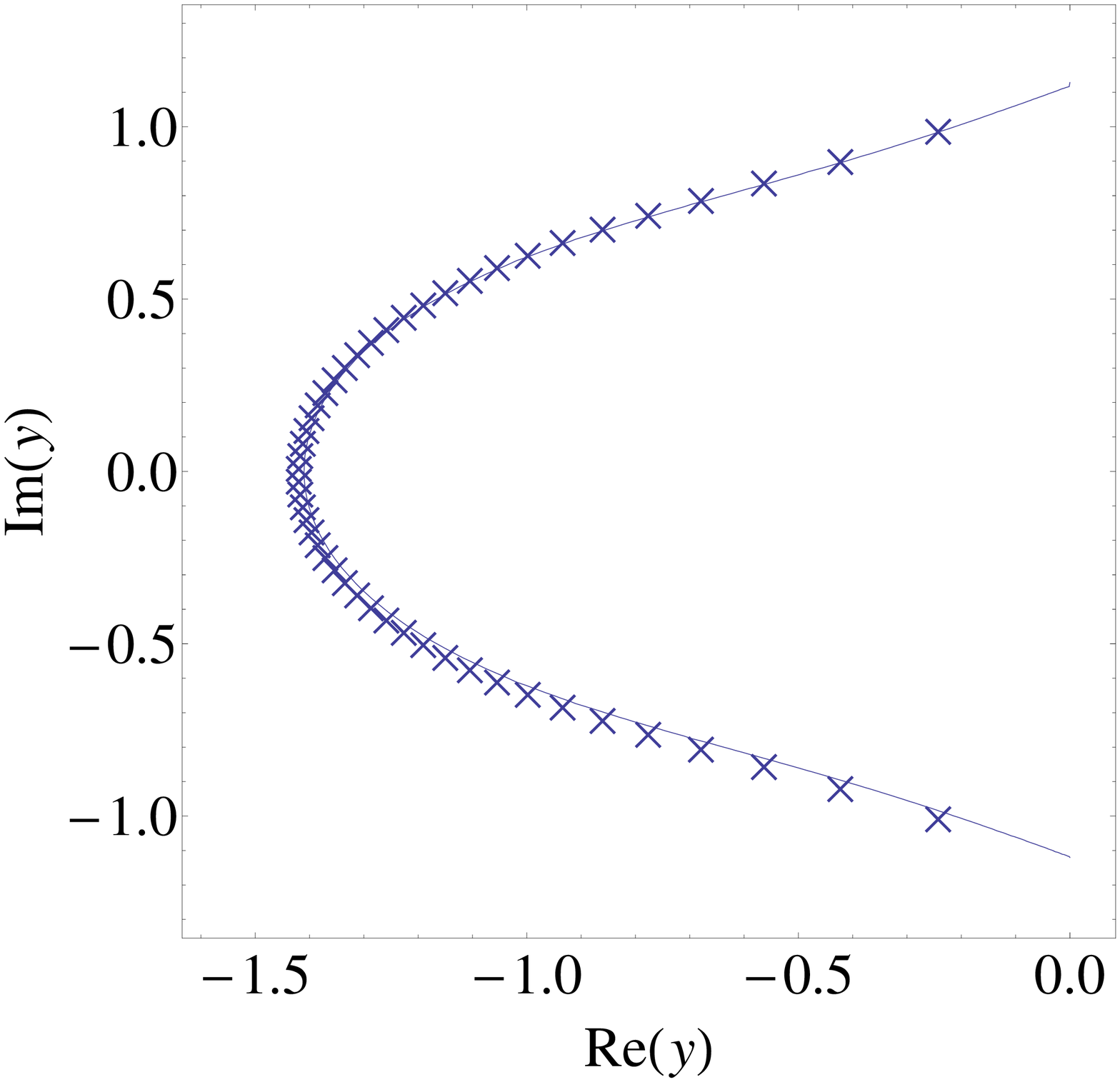}}
\caption{Roots for the ground state of the degenerate, two-level Richardson model with $\varepsilon_{2}=1$, $\varepsilon_{1}=-1$, $L=100$, and $M=50$: (a) $g={1}/{2}$, (b) $g=1$, and (c) $g={3}/{2}$. Also shown are the theoretical curves derived in Appendix B for the  
large-$L$ limit. }
\label{fig1}
\end{figure}

\begin{figure}
\centering
\subfloat[]{\includegraphics[width=5cm]{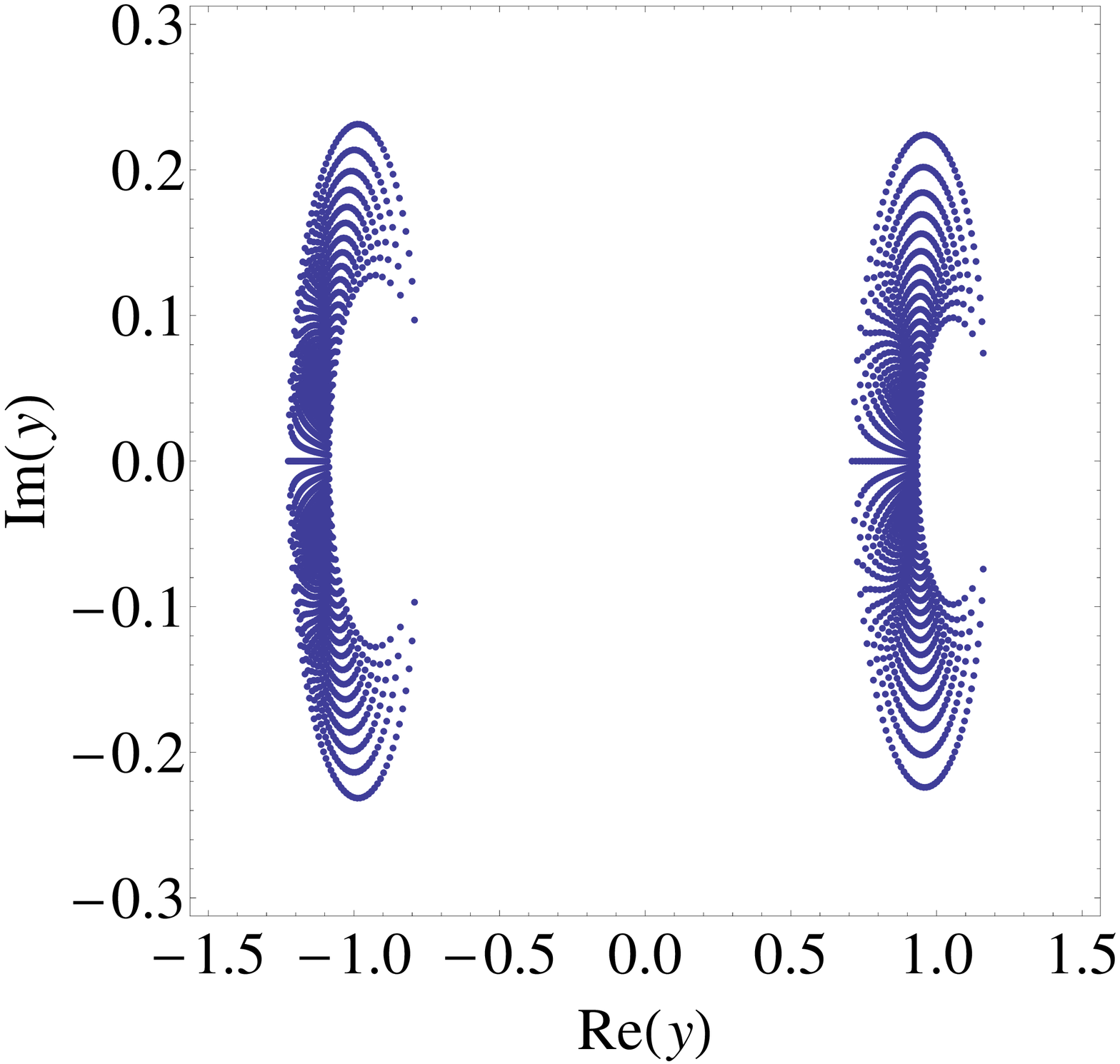}}
\subfloat[]{\includegraphics[width=5cm]{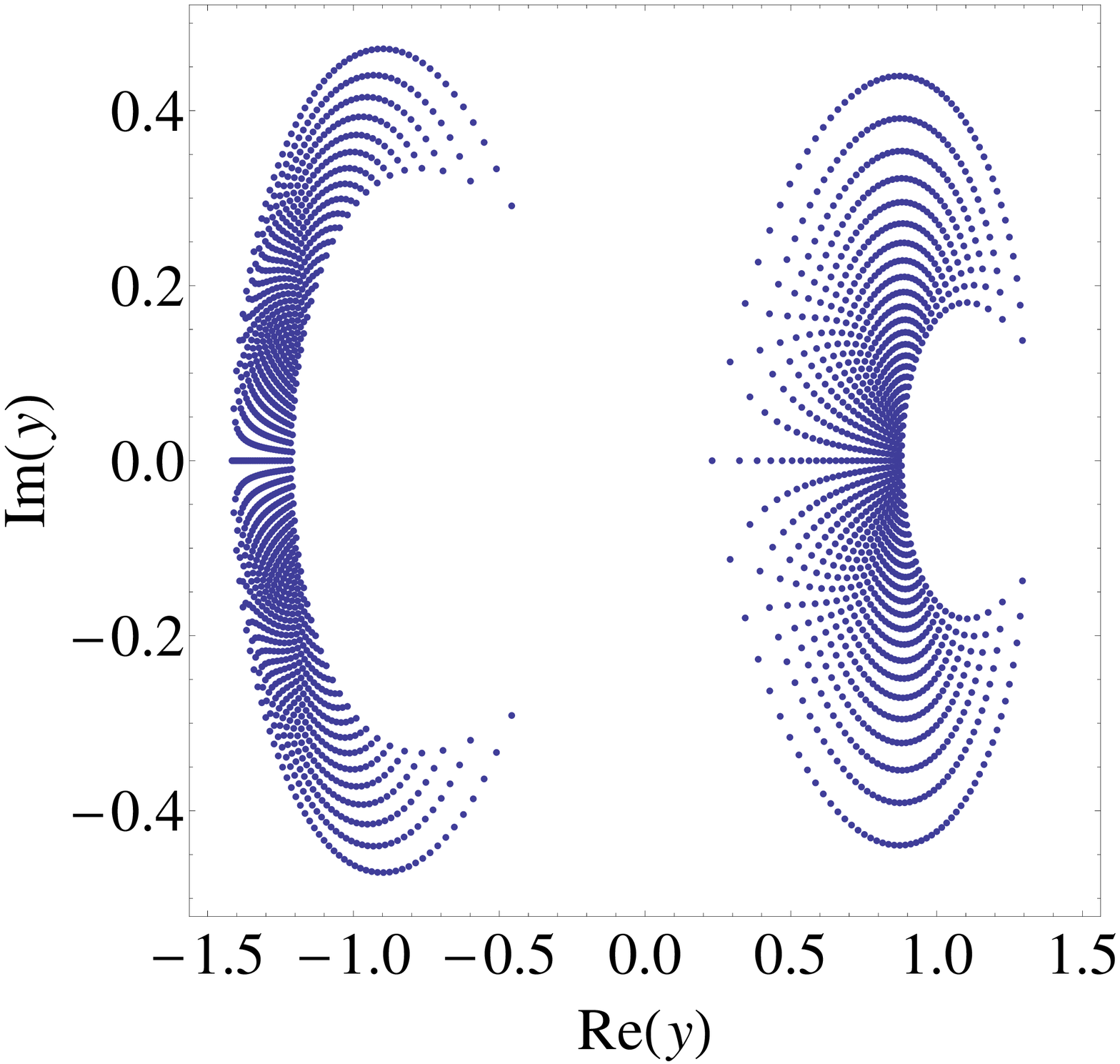}}
\subfloat[]{\includegraphics[width=5cm]{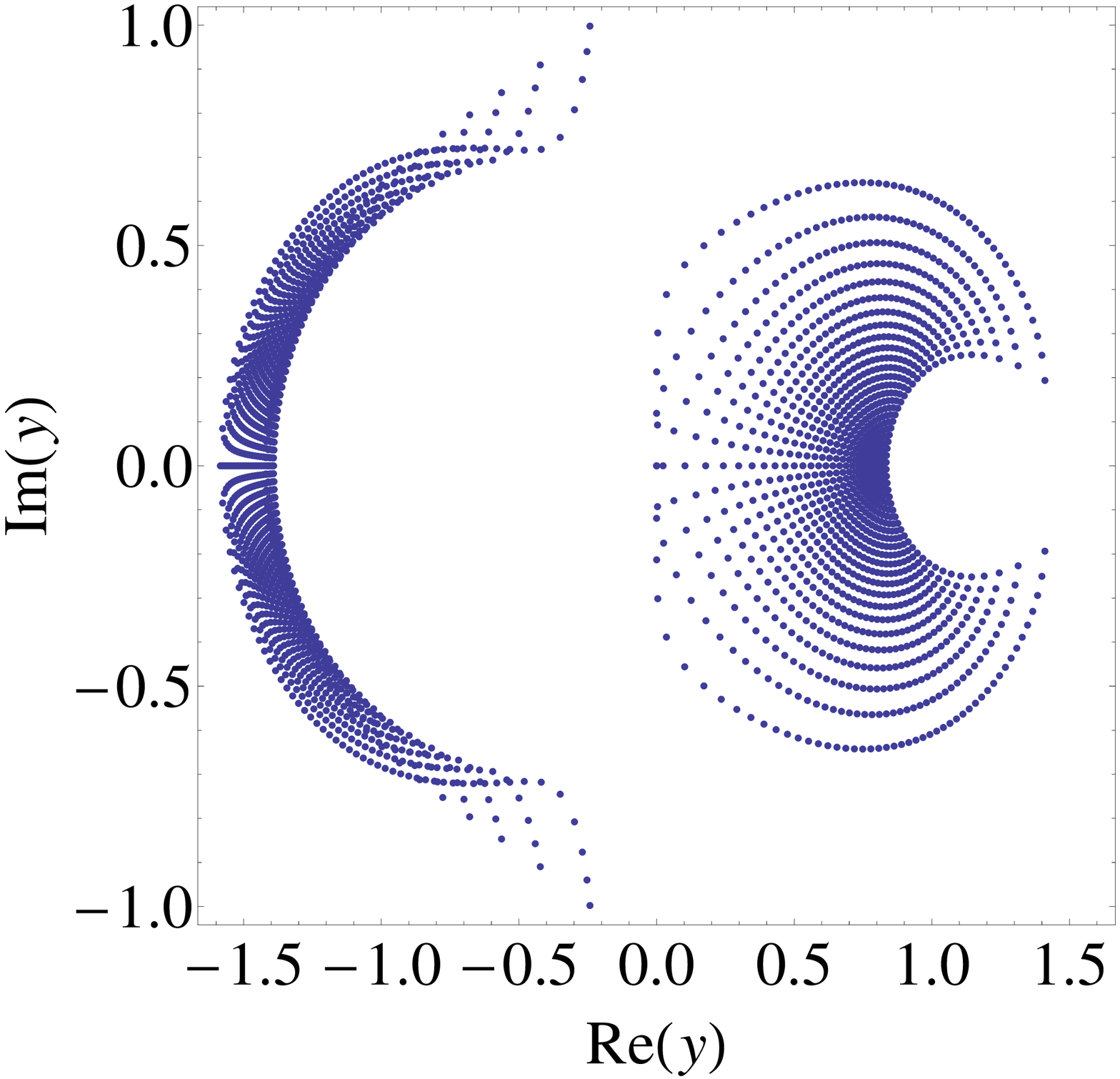}}
\caption{Roots for all unblocked symmetric states of the degenerate, two-level $s$-wave pairing model with $\varepsilon_{2}=1$, $\varepsilon_{1}=-1$, $L=100$, and $M=50$: (a) $g={1}/{2}$, (b) $g=1$, and (c) $g={3}/{2}$. In each case a total of 51 sets of roots are displayed, where each set contains 50 roots.}
\label{fig1}
\end{figure}

\subsection{\textbf{Example 2: The $p+ip$-wave pairing Hamiltonian}}

Next we examine the $p+ip$-wave pairing model, whose exact solution has only recently been derived \cite{Iba,Dun2}. The connection to Heine-Stieltjes and Van Vleck polynomials for this model was noted in \cite{Romb}. Up to a canonical transformation, the Hamiltonian reads 
\begin{align*}
H=\sum_{i=1}^{L}\varepsilon_{i}N_{i}-G\sum_{j<k}^{L}\sqrt{\varepsilon_{j}\varepsilon_{k}}(b_{j}^{\dagger}b_{k}+b_{k}^{\dagger}b_{j}). %\label{eq19}
\end{align*}
A notable feaure of this model is the ground-state phase diagram, which is summarised in Fig. 2. The phase boundaries known as the Moore-Read line and the Read-Green line have the property that they are independent of the parameters $\varepsilon_j$. Consequently, signatures of the phase boundaries should still be present in the degenerate two-level limit.

The energy of each eigenstate is given as the sum
\begin{equation}
E=(1+G)\sum_{j=1}^{M}y_{j}, \label{eq21}
\end{equation}
where the roots $y_{j}$ satisfy Bethe Ansatz equations of the form given by Eq. (\ref{eq2}) with $B={G}^{-1}-L+2M-1$  and $A=C=0$. It thus belongs to case 6 of Table 1.
In the degenerate two-level limit, the corresponding second order differential equation is 
\begin{align*}
(z^{3}-\gamma z^{2}+\eta z)Q''+\left(\frac{\eta (-1+G (1+L-2 M))}{G}+\frac{\gamma (2-G (2+L-4 M)) z}{2 G}\right. %\label{eq22}
\end{align*}
\[+\left.\frac{(-1+G-2 G M) z^{2}}{G}\right)Q'- (\beta_{1}z +\beta_{0})Q=0.\]
Using the expansion of the polynomial $Q(z)$ and the energy expression (\ref{eq21}) leads to
\begin{align*}
\beta_{1}=-\left(\frac{M}{G}+M^{2}\right),\quad \beta_{0}=-\left(\frac{E}{G}-\frac{\gamma M}{G}+\frac{\gamma L M}{2}-\gamma M^{2}\right). %\label{eq23}
\end{align*}
\begin{table}
\caption{Phase diagram of the $p+ip$-wave pairing model}
\footnotesize\rm
\begin{tabular*}{\textwidth}{@{}l*{15}{@{\extracolsep{0pt plus12pt}}l}}
\br
  & Phase   &   Constraint between coupling constant $g$ and filling fraction $x$    \\
\br
  1 & Weak coupling BCS  & $ x > 1-g^{-1}$    \\
  2 & Moore-Read line  & $x_{MR}=1-g^{-1}$  \\
  3 & Weak pairing  & $({1-g^{-1}})/{2} < x < 1-g^{-1}$ \\
  4 & Read-Green line  & $x_{RG}=({1-g^{-1}})/{2}$ \\
  5 & Strong pairing   & $ x <  {(1-g^{-1})}/{2}$  \\
  
\mr
\end{tabular*}
\end{table}

 \begin{figure}[ht!]
    \label{fig:subfigures}
    \begin{center}
        \subfloat[]{  \label{fig:first}  \includegraphics[width=0.4\textwidth]{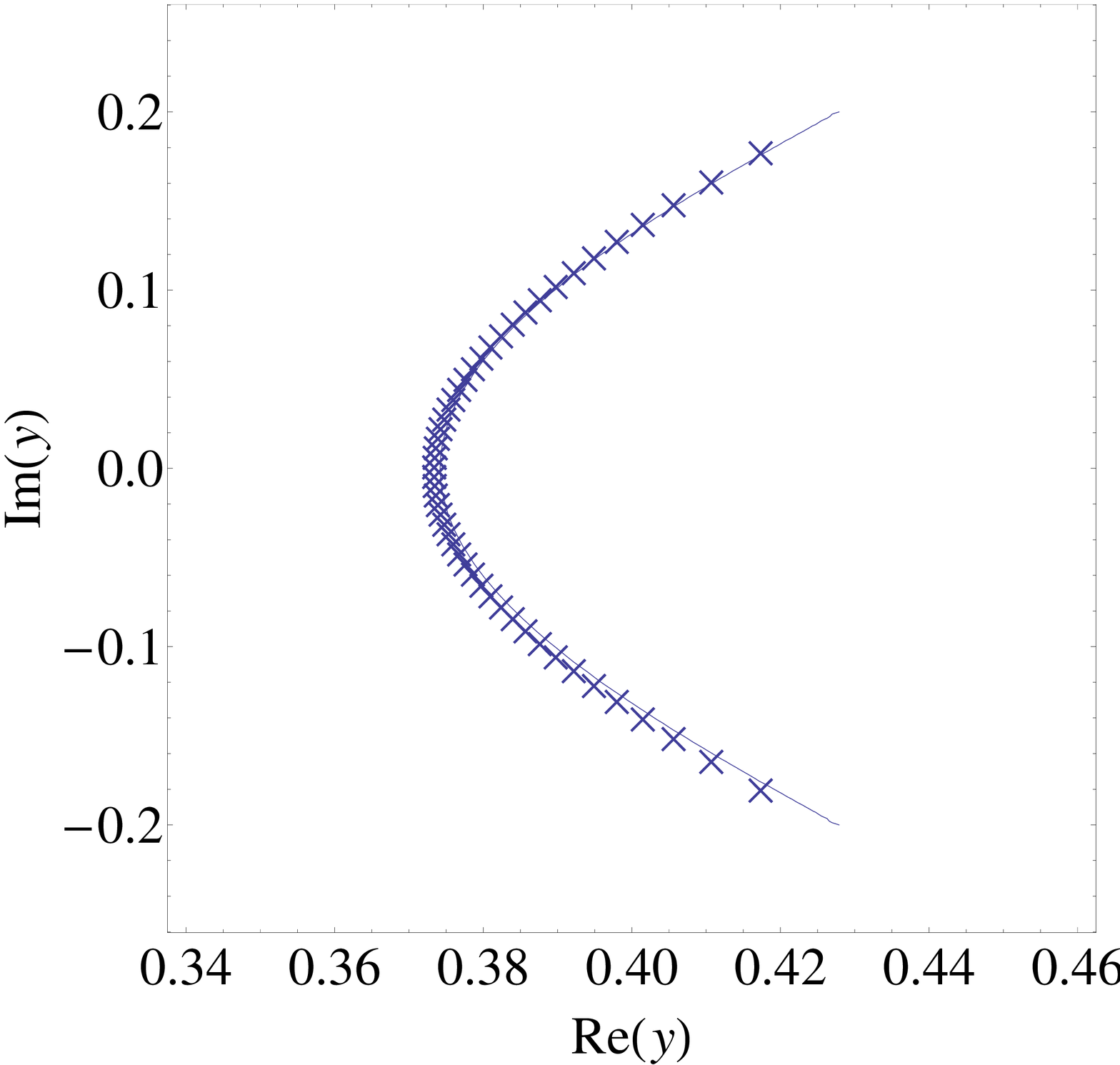}  }
        \subfloat[]{  \label{fig:second} \includegraphics[width=0.4\textwidth]{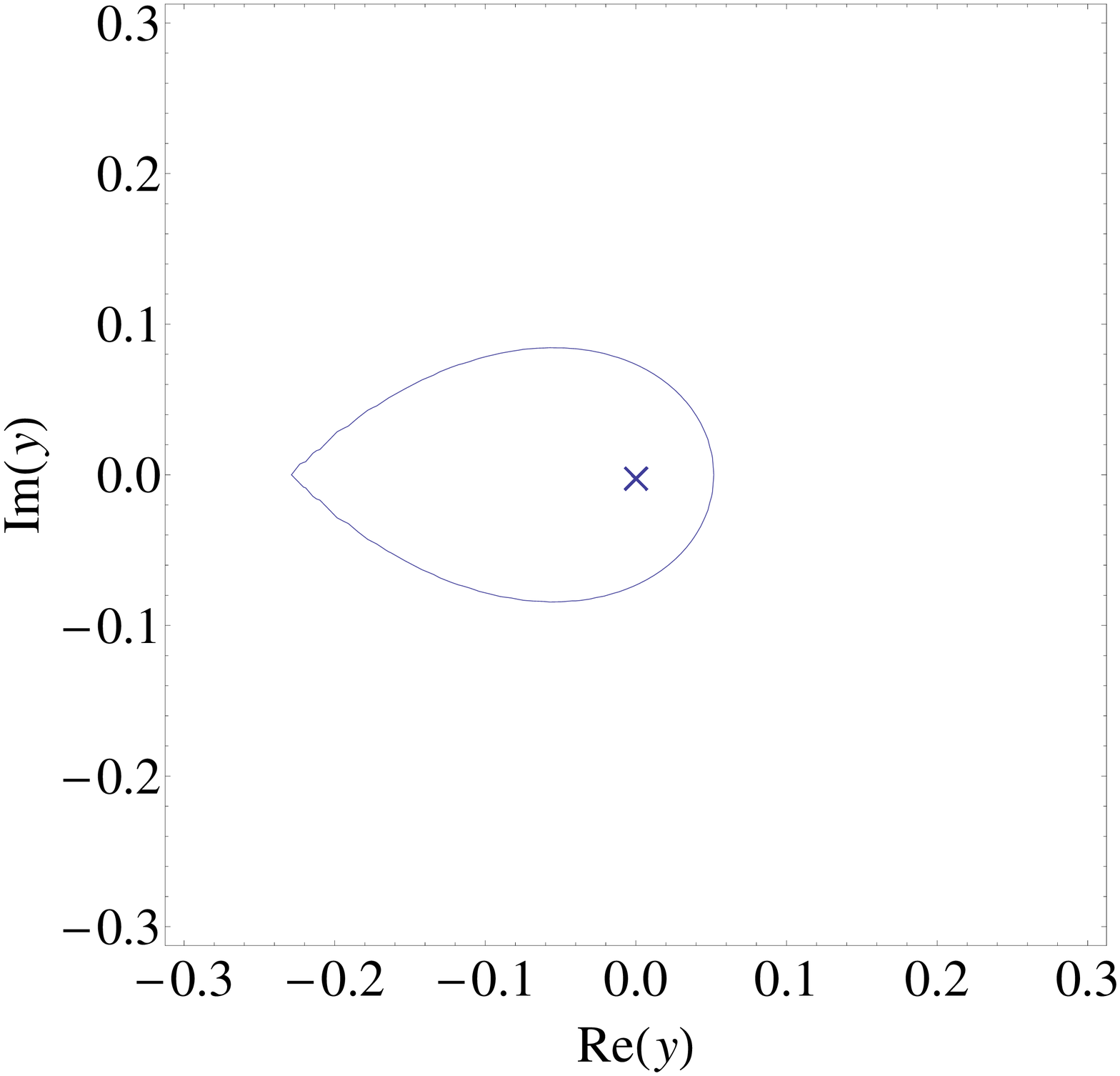} }\\ 
        \subfloat[]{  \label{fig:third}  \includegraphics[width=0.4\textwidth]{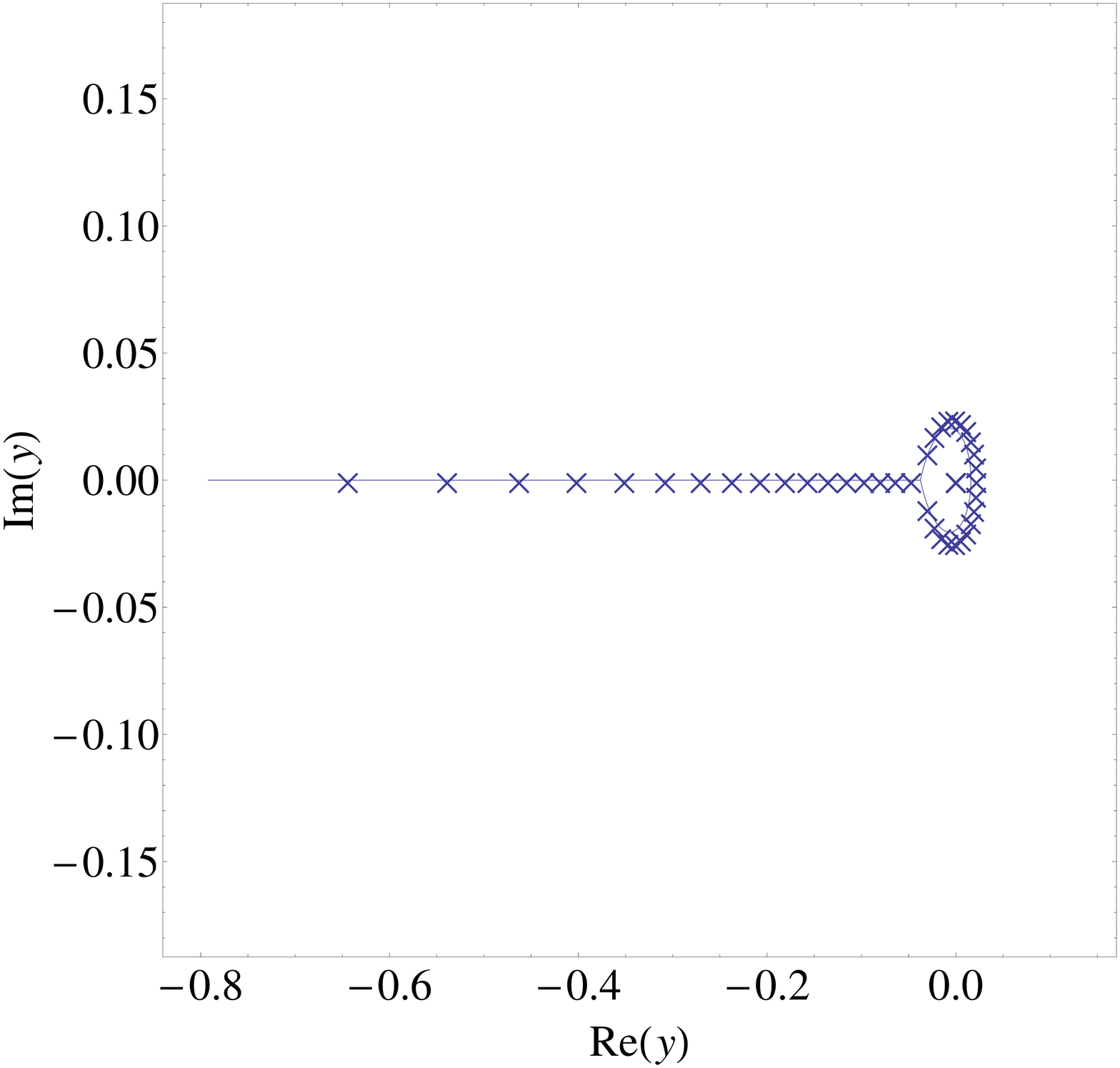} }
        \subfloat[]{  \label{fig:fourth} \includegraphics[width=0.4\textwidth]{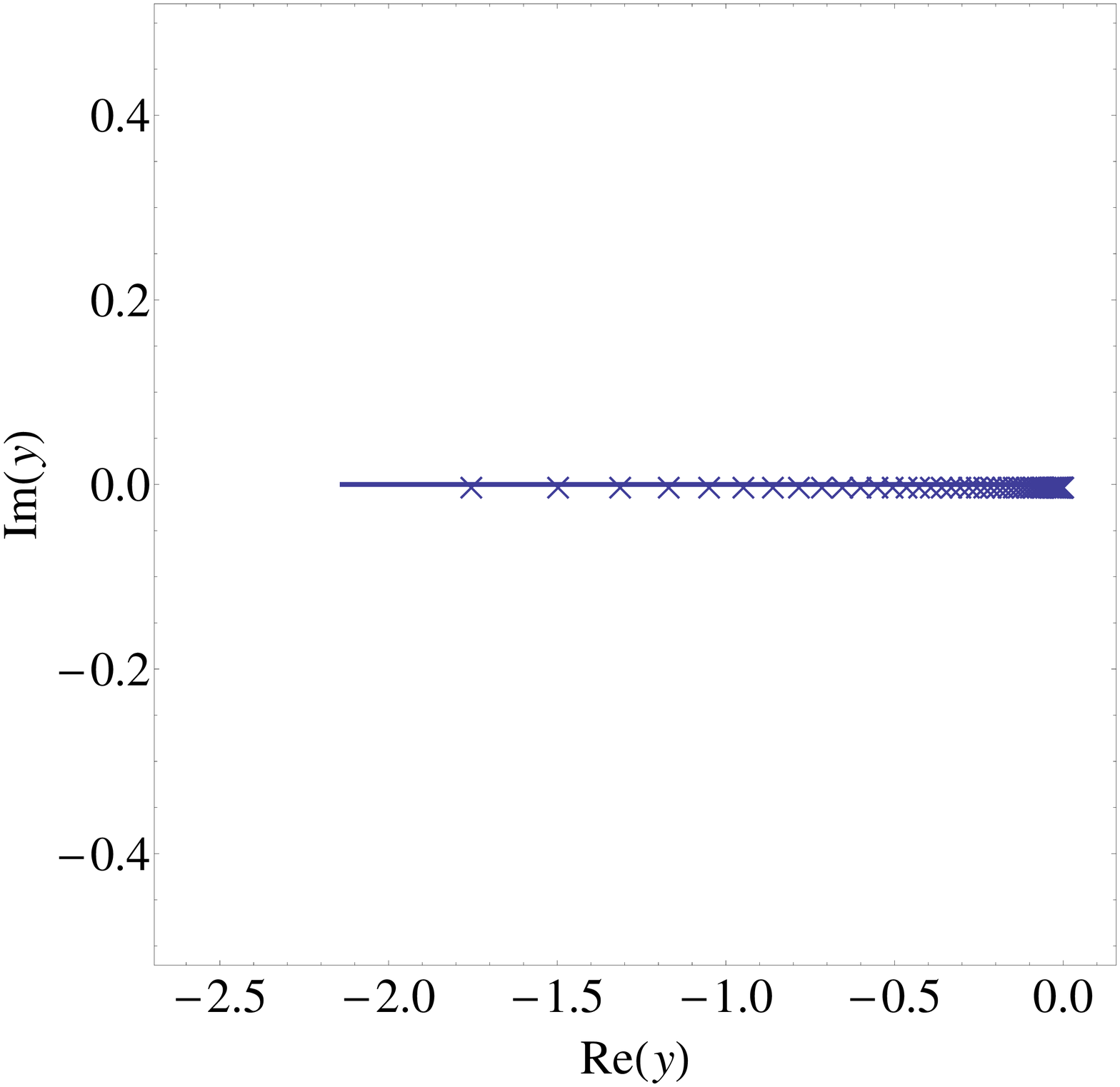} }
    \end{center}
    \caption{%
          Roots for the ground state of the degenerate, two-level $p+ip$-wave pairing model with $\varepsilon_{2}=1$ and $\varepsilon_{1}={1}/{2}$, $L=200$, and $M=50$: (a) $g={1}/{2}$ (Weak coupling BCS), (b) $g={4}/{3}$ (Moore-Read line), (c) $g={3}/{2}$ (Weak pairing), and (d) $g=2$ (Read-Green line). Also shown are the theoretical curves derived in Appendix B for the  large-$L$ limit.  
     }%
\end{figure}

\begin{figure}[ht!]
    \label{fig:subfigures}
    \begin{center}
        \subfloat[]{\label{fig:first} \includegraphics[width=6cm]{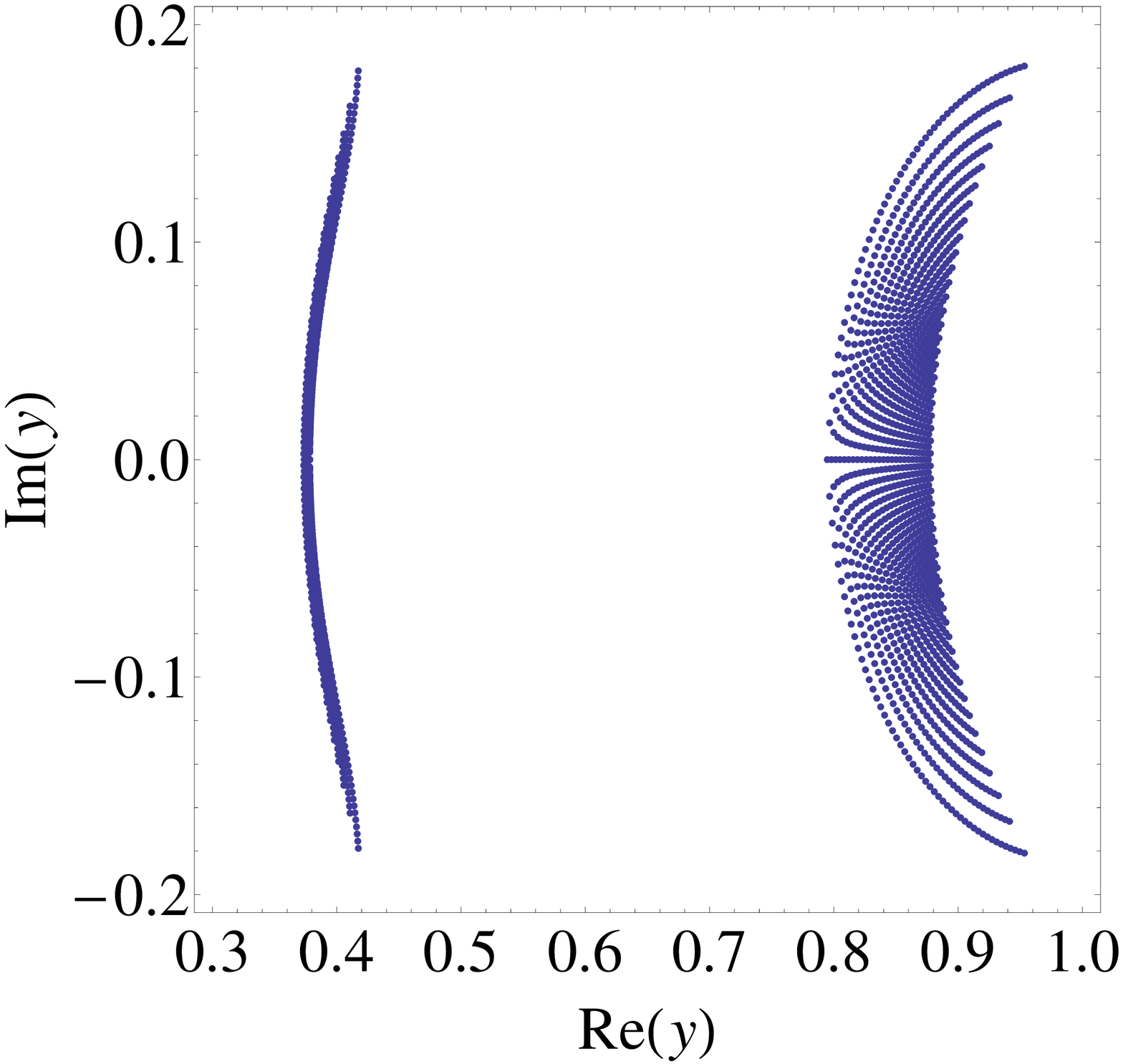}}
        \subfloat[]{ \label{fig:second} \includegraphics[width=6cm]{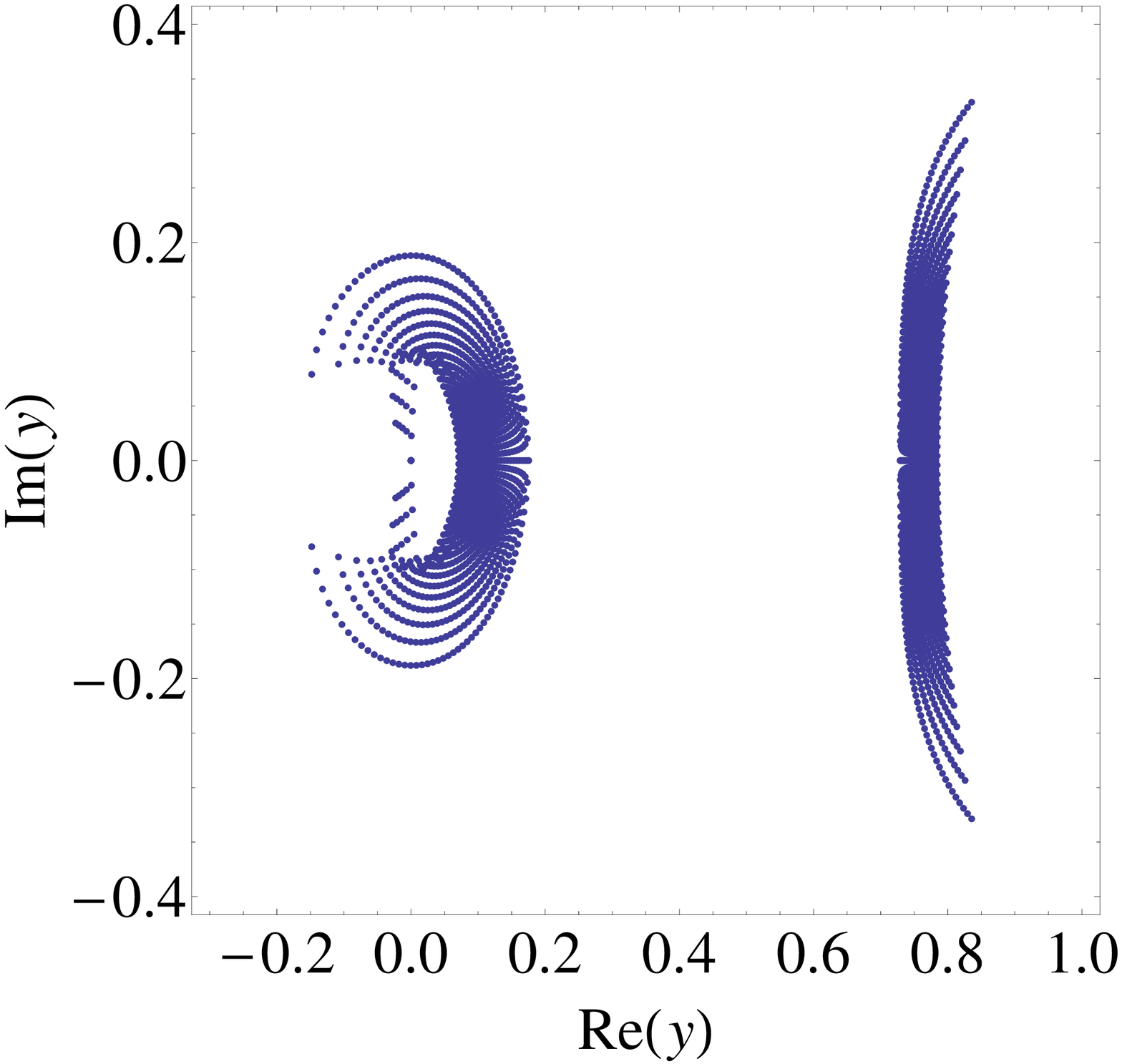}} \\  
        \subfloat[]{ \label{fig:third} \includegraphics[width=6cm]{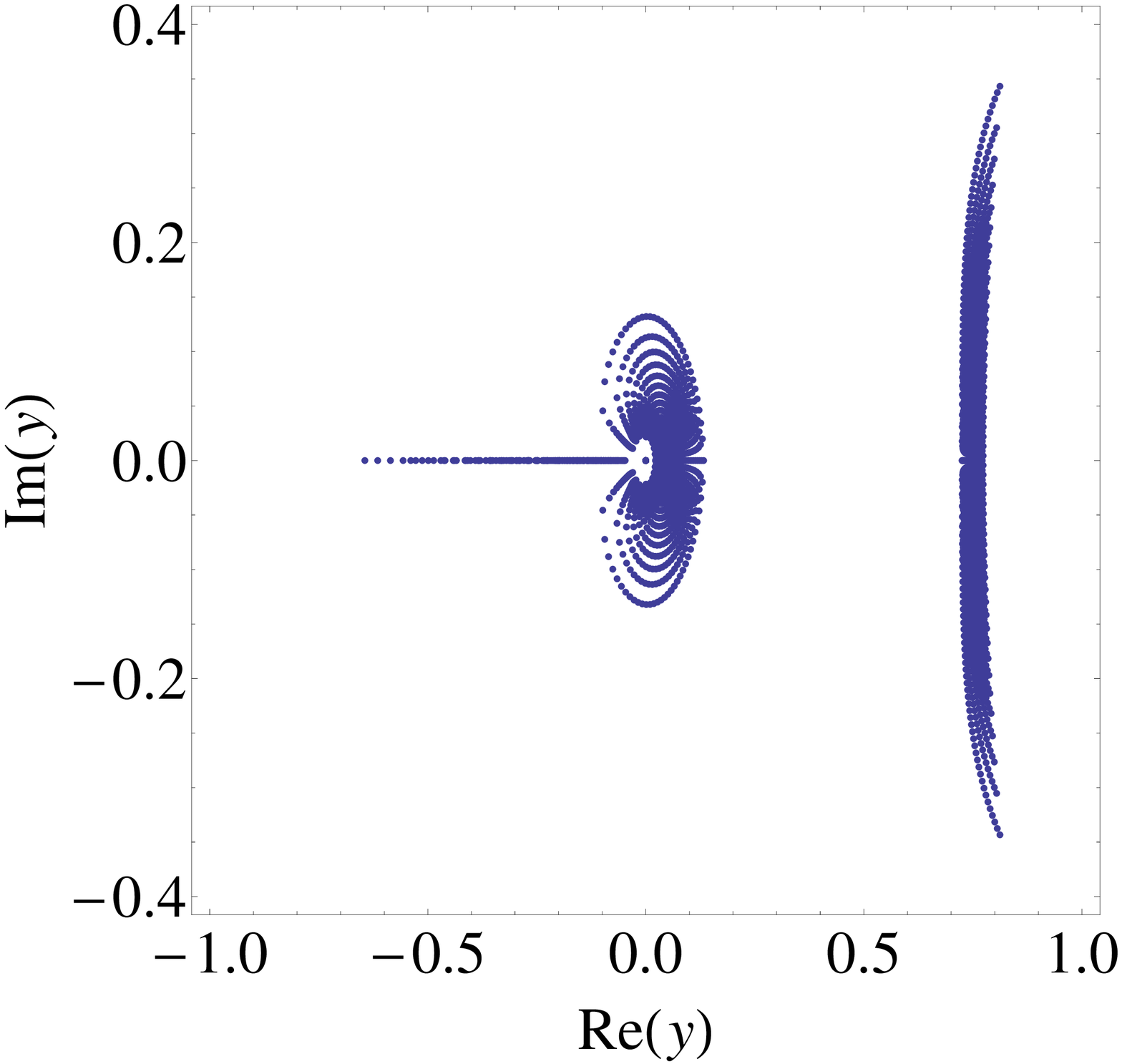}  } 
        \subfloat[]{  \label{fig:fourth} \includegraphics[width=6cm]{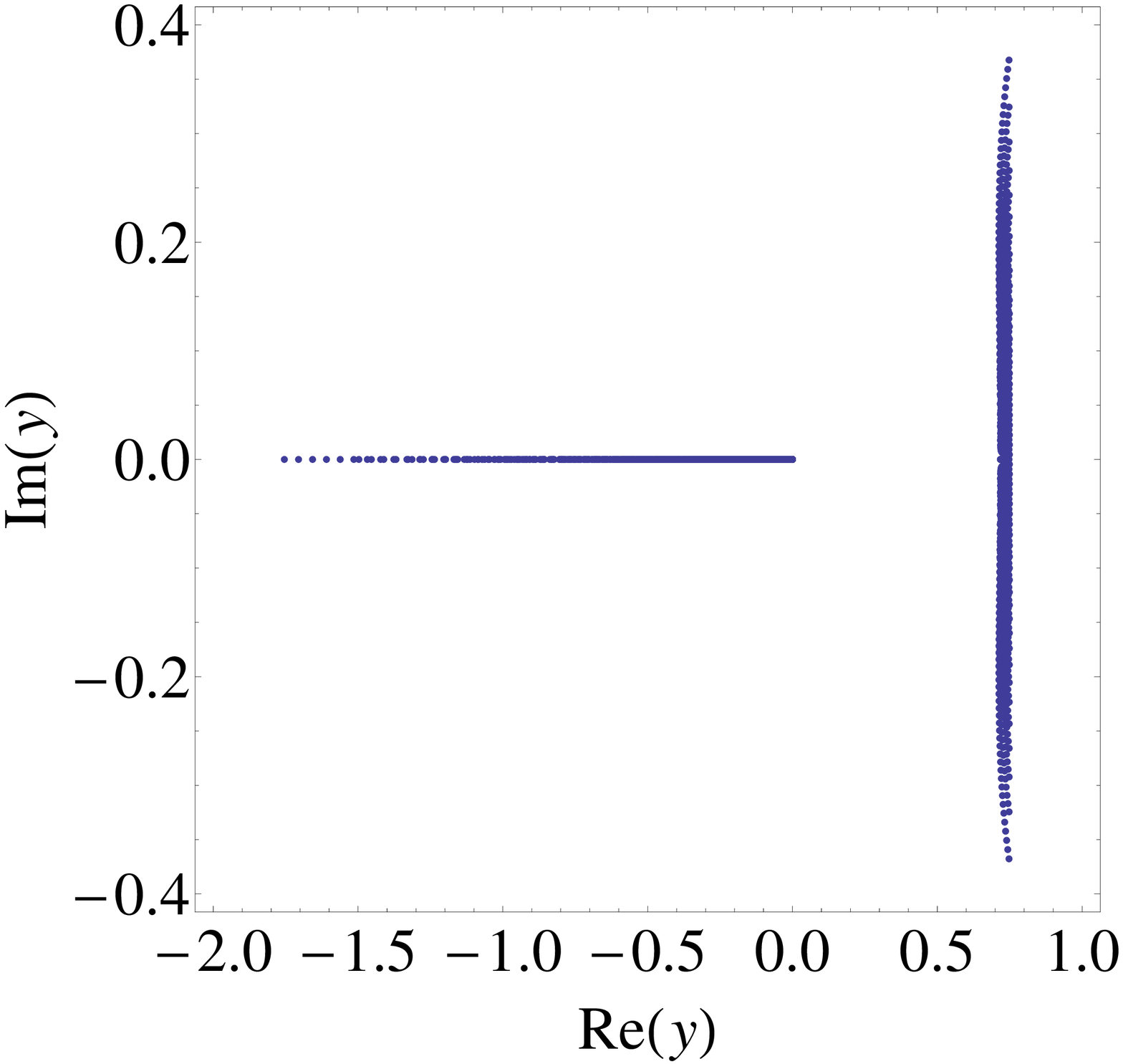} }
    \end{center}
    \caption{Roots for all unblocked symmetric states of the degenerate, two-level $p+ip$-wave pairing model with $\varepsilon_{2}=1$ and $\varepsilon_{1}={1}/{2}$, $L=200$, and $M=50$: (a) $g={1}/{2}$, (b) $g={4}/{3}$, (c) $g={3}/{2}$, and (d) $g=2$. In each case a total of 51 sets of roots are displayed, where each set contains 50 roots.}
\end{figure}

We consider the case $\varepsilon_{2}=1$, $\varepsilon_{1}={1}/{2}$, $L=200$, and $M=50$.  Ground state roots are depicted in Fig. 3, along with the theoretical curves derived in Appendix B for the  large-$L$ limit.  A curious feature here is case (b) corresponding to the Moore-Read line. At this point there is a change from an open curve to a closed curve in the continuum limit, however all the roots collapse to the origin in any finite system.  This is a result of the limits $g\rightarrow g_{MR}$ and $L\rightarrow\infty$ not commuting, and was identified in \cite{Iba,Dun2} as a zeroth order quantum phase transition. This is in stark contrast to Fig. 1 (b). We also highlight that for case (c) the curve consists of two connected components, one of which is a closed curve and another which is open. At the Read-Green line shown in (d) the closed curve has contracted to a point at the origin. The theoretical curves shown in Fig. 3 have exactly the same qualitative features as is found in the general $p+ip$-wave pairing model \cite{Dun2,Romb}.     
As in the previous example, the number of states in the $M=50$ sector is 51, so the number of data points in each panel of Fig. 4 is $2550$.  

\subsection{\textbf{Example 3: The $p+ip$-wave pairing Hamiltonian coupled to a bosonic molecular pair degree of freedom}}

We next consider the case of the  $p+ip$-wave pairing Hamiltonian coupled to bosonic molecular pair \cite{Dun3}. This model is given by the following Hamiltonian
\begin{align*}
H=\sum_{j=1}^{L}\varepsilon_{j}N_{j}-F^{2}GN_{0}-G\sum_{j<k}^{L}\sqrt{\varepsilon_{j}\varepsilon_{k}}(b_{j}^{\dagger}b_{k}+b_{k}^{\dagger}b_{j}) -FG\sum_{j=1}^{L}\sqrt{\varepsilon_{i}}(b_0b_{j}^{\dagger}+b_0^{\dagger}b_{j})  %\label{eq24}
\end{align*}
where $b_0,\,b_0^\dagger$ satisfy the bosonic commutation relation
$$[b_0,\,b_0^\dagger]=I$$
and $N_0=b_0^\dagger b_0$. 
This system has the energy spectrum
\begin{align*}
E=(1+G)\sum_{j=1}^{M}y_{j}, %\label{eq25}
\end{align*}
where the roots $y_{j}$ satisfy the Eq.(\ref{eq2}) with $C=0$, $B={G}^{-1}+2M-L-1$ and  $A=F^2 $, thus belonging to case 2. 
We form the following differential equation 
\begin{align*}
(z^{4}-\gamma z^{3}+\eta z^{2})Q''+(-F^{2} \eta+(F^{2} \gamma+\eta-\frac{\eta}{G}+\eta L-2 \eta M) z %\label{eq26}
\end{align*}
\[ +(-F^{2}-\gamma+\frac{\gamma}{G}-\frac{\gamma L}{2}+2 \gamma M) z^{2}+(1-\frac{1}{G}-2 M) z^{3})Q'-(\beta_{2}z^{2}+\beta_{1}z+\beta_{0})Q=0.\]
The coefficient of the generalised Van Vleck polynomial can be rewritten in the following form
\begin{align*}
\beta_{2}=-M \left(\frac{1}{G}+M \right),\quad \beta_{1}=-\left(\frac{2E+(-2(\gamma-F^{2} G)+\gamma G L) M-2 \gamma G M^{2}}{2 G}\right).  %\label{eq27}
\end{align*}
In this instance, some coefficients of $A_{1}$ and all coefficients of $A_{0}$ depend of $M$. The coefficient $\beta_{0}$ is more complicated than for the previous examples, as it involves a double sum of the roots and is not given only in terms of the energy. 

We consider the choice $\varepsilon_{2}=1$ and $\varepsilon_{1}={1}/{2}$, $L=32$, $M={L}/{2}$ and $F=\sqrt{128}$ to undertake the numerical calculations. Ground-state roots are shown in Fig. 5. They always lie on the negative real-axis. The roots of all unblocked symmetric states are shown in Fig. 6. There are no significant qualitative changes in the pattern of roots as the coupling parameter $g$ is varied.

\begin{figure}
\centering
\subfloat[]{\includegraphics[width=5.2cm]{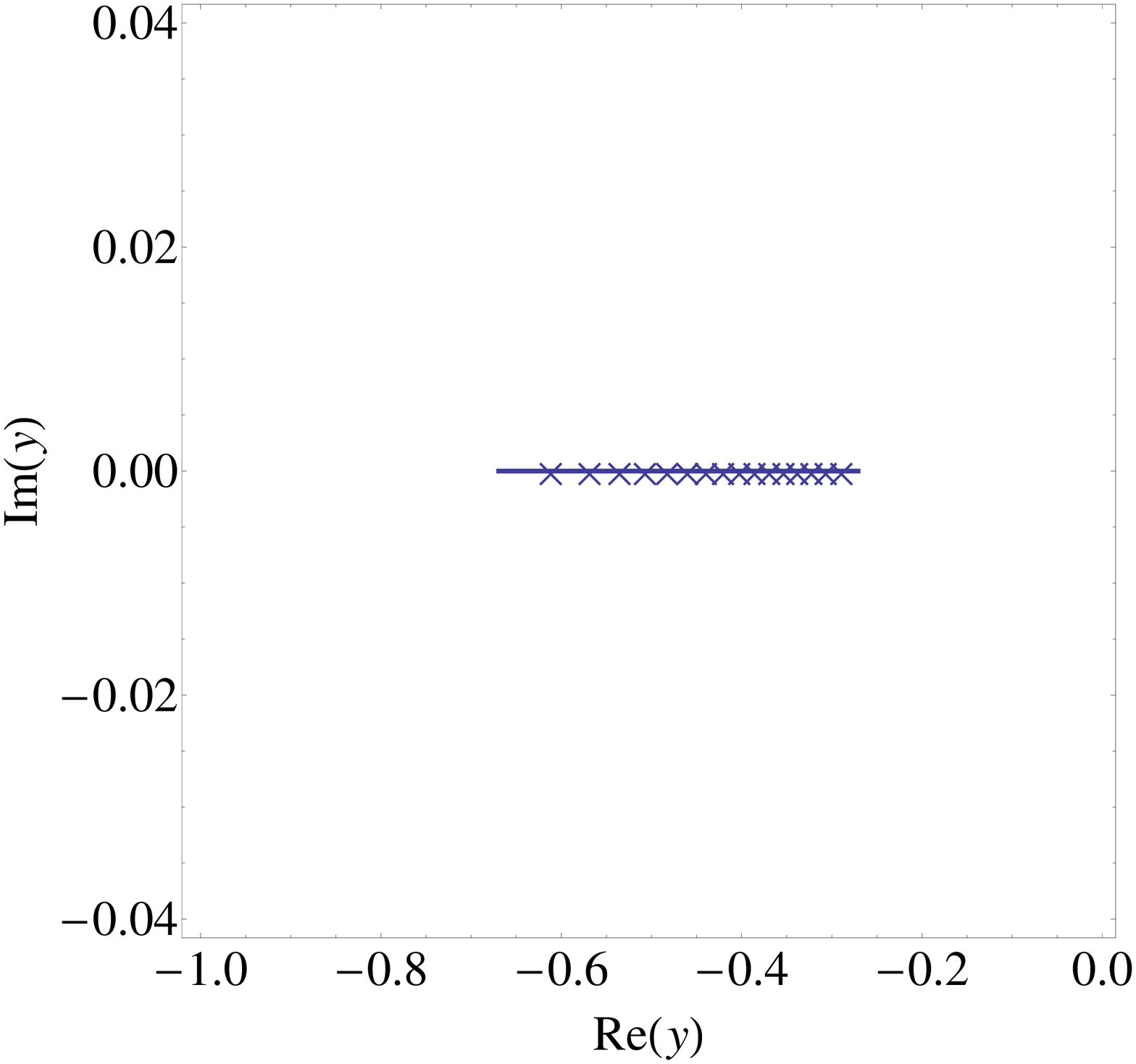}}
\subfloat[]{\includegraphics[width=5cm]{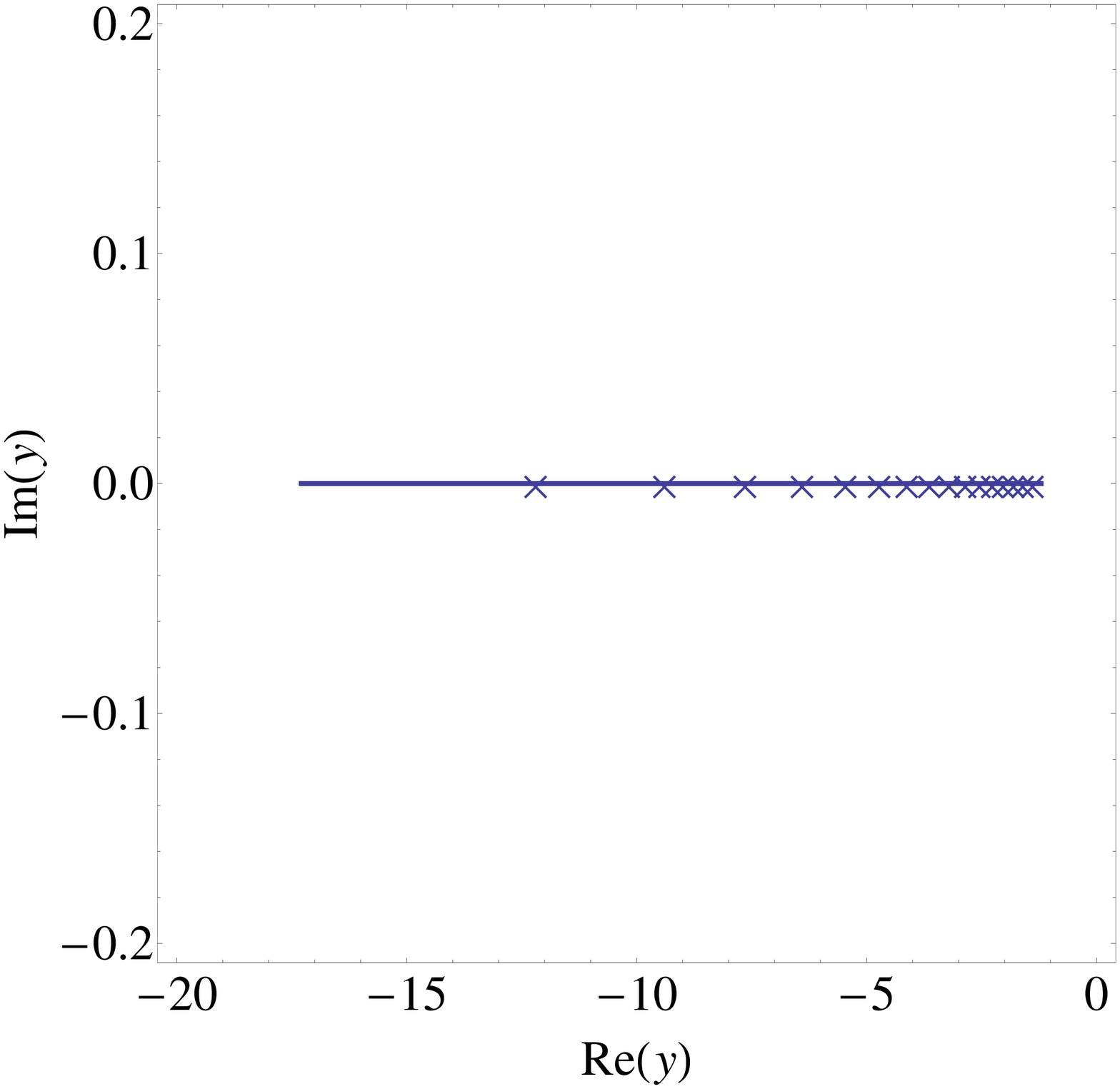}}
\subfloat[]{\includegraphics[width=5cm]{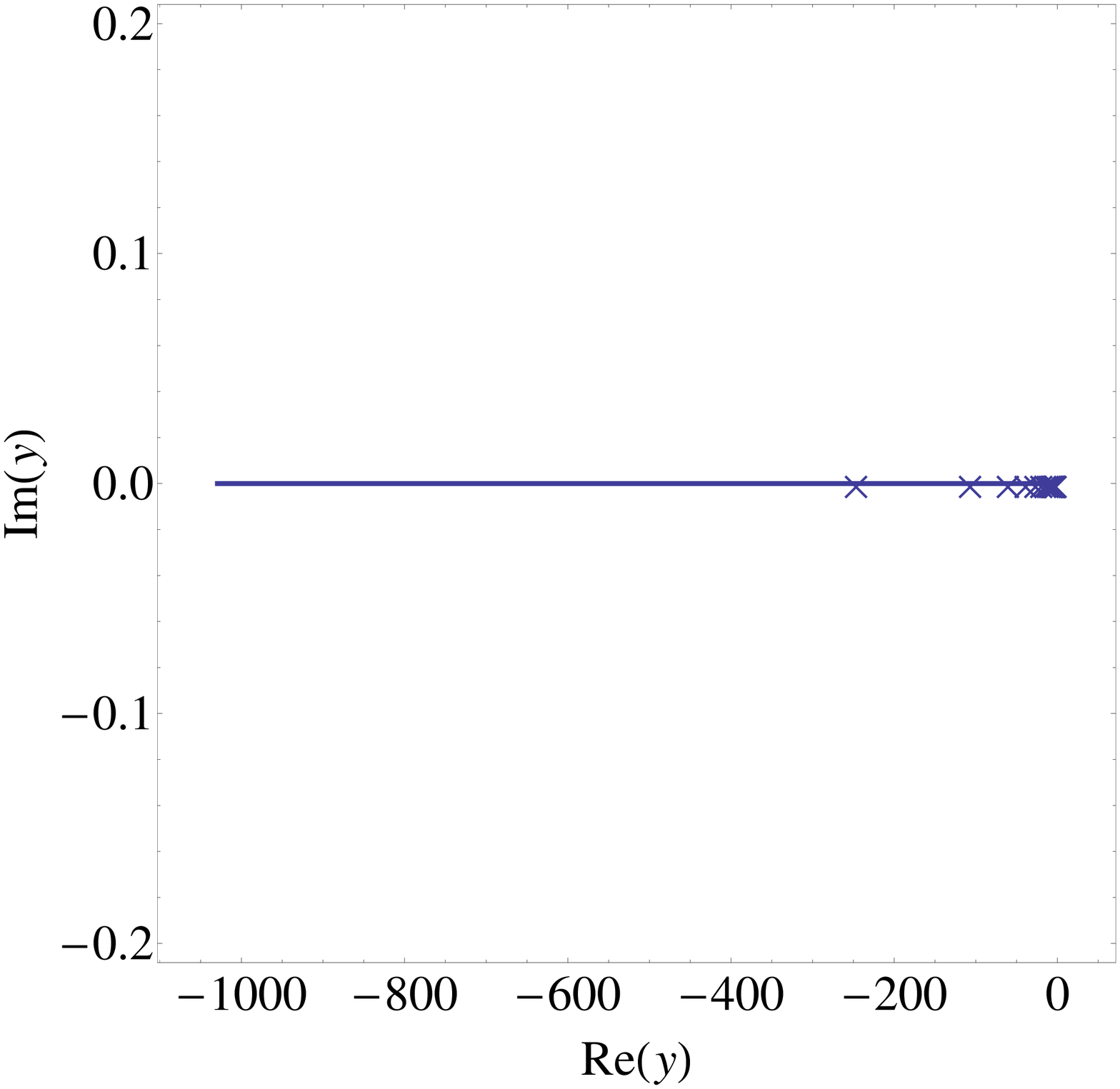}}
\caption{Roots for the ground state of the degenerate, two-level $p+ip$-wave pairing model coupled to a bosonic molecular pair with $\varepsilon_{2}=1$,  $\varepsilon_{1}={1}/{2}$, $L=32$, $M=16$, and $F=\sqrt{128}$: (a) $g={1}/{10}$, (b) $g=1$, and (c) $g=10$. Also shown are the theoretical curves derived in Appendix B for the  large-$L$ limit. }
\label{fig1}
\end{figure}

\begin{figure}
\centering
\subfloat[]{\includegraphics[width=5cm]{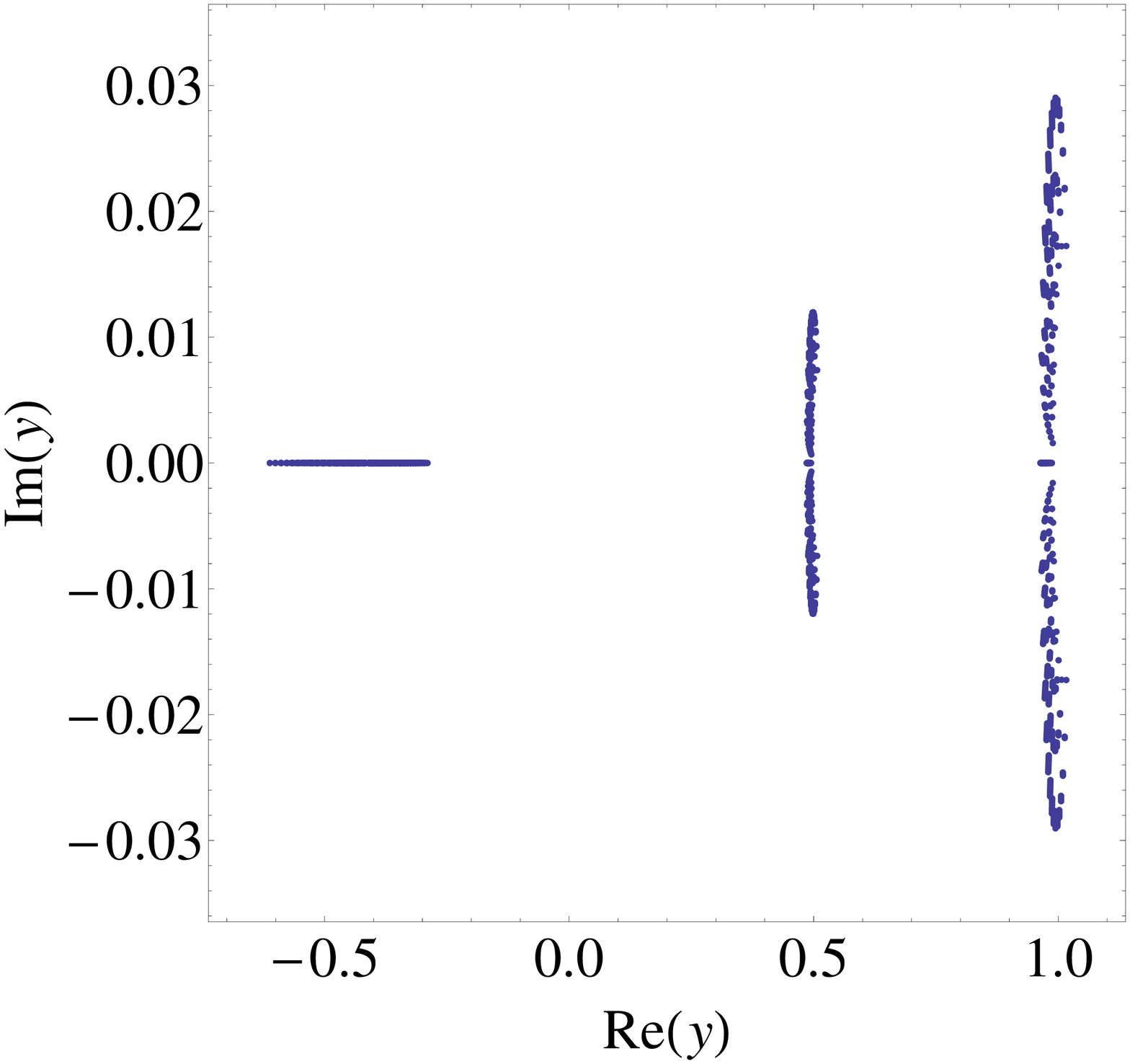}}
\subfloat[]{\includegraphics[width=5cm]{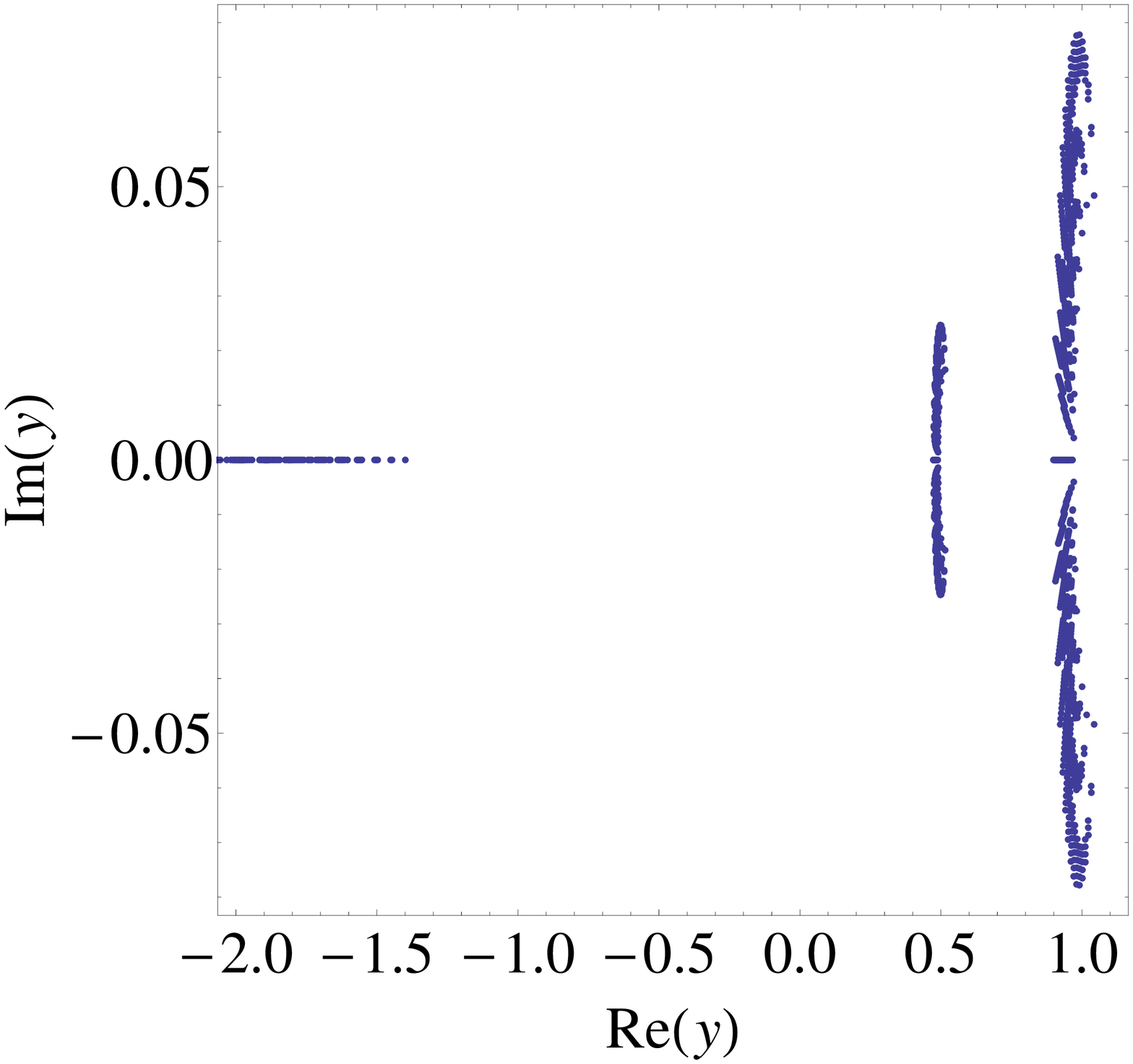}}
\subfloat[]{\includegraphics[width=5cm]{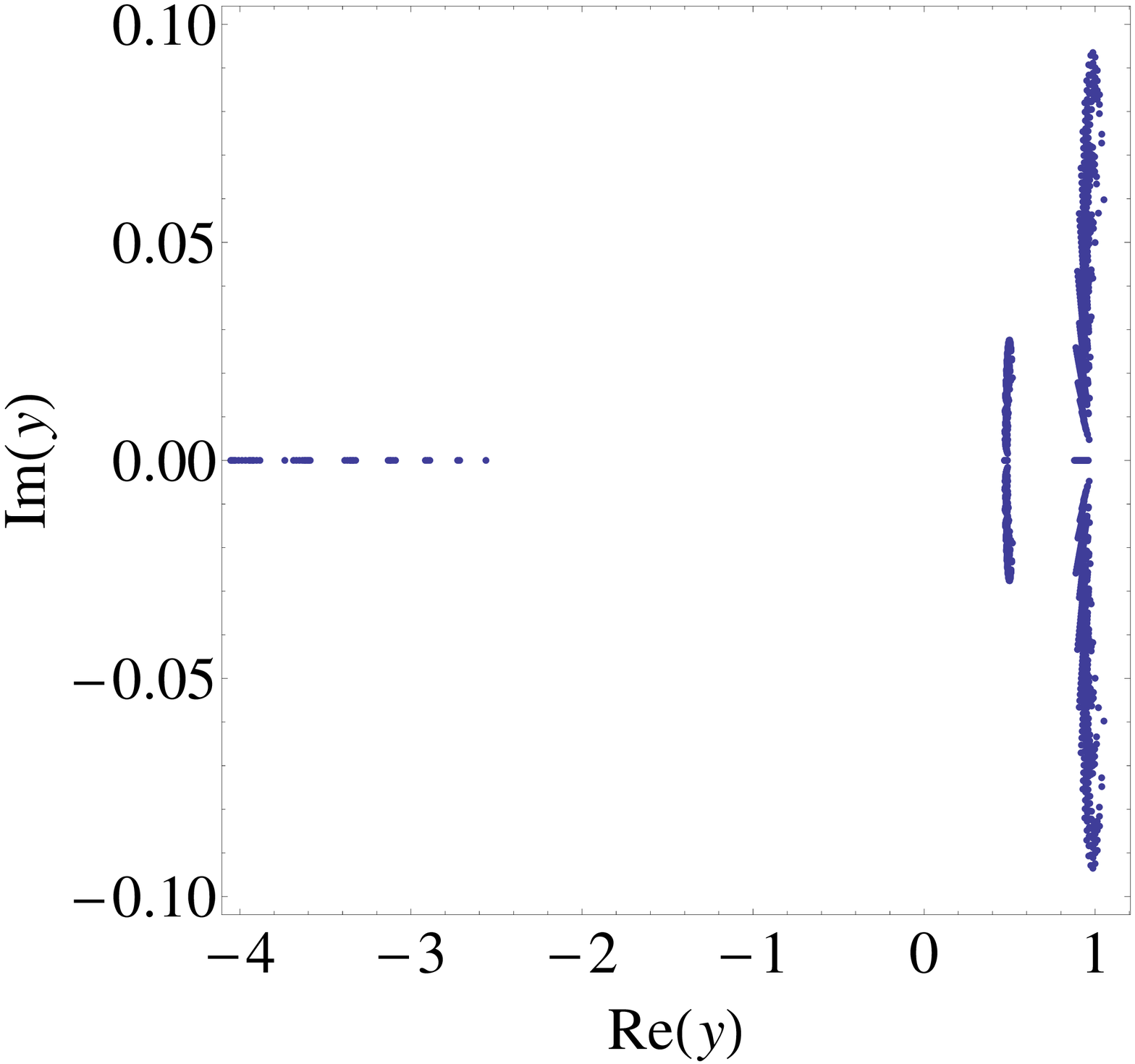}}
\caption{Roots for all unblocked symmetric states of the degenerate, two-level  $p+ip$-waved pairing model coupled to a bosonic molecular pair with $\varepsilon_{2}=1$, $\varepsilon_{1}={1}/{2}$, $L=32$, $M=16$, and $F=\sqrt{128}$: (a) $g={1}/{10}$, (b) $g=1$, (c) $g=10$.    In each case a total of 153 sets of roots are displayed, where each set contains 16 roots. }
\label{fig1}
\end{figure}

\subsection{\textbf{Example 4: An extended $d+id$-pairing Hamiltonian}}

Using the results concerning the conserved operators of the Richardson model, it is possible through a change of variables for the single particle levels to obtain an integrable Hamiltonian given by \cite{ml12}
\begin{align*}
H=\sum_{i=1}^{L}\varepsilon_{i}N_{i}-G\sum_{j,k=1}^{L}\varepsilon_{j}\varepsilon_{k}(b_{j}^{\dagger}b_{k}+b_{j}b_{k}^{\dagger}+2N_{j}N_{k}). %\label{eq28}
\end{align*}
In the absence of the $N_j N_k$ interaction terms, this model maps through a canonical transformation to a $d+id$-pairing Hamiltonian. 
%The conserved quantities :
%\begin{equation}
%\tau_{j}=(\frac{1}{2G}-\sum_{j}k_{j})\frac{1}{2}(N_{j}-I)+\sum_{k\neq j}^{L}\frac{\theta_{jk}k_{j}k_{k}}{k_{k}-k_{j}}, \label{eq29}
%\end{equation}

%with $\theta$ given by Eq.(\ref{eq14})

%and the 
The energies are given by 
\begin{align*}
E=\sum_{l=1}^{M} y_{l}-2 G \sum_{l=1}^{M}\sum_{j\neq l}^{M} y_{j}y_{l}-G \sum_{j=1}^{L}\varepsilon_{j}. %\label{eq30}
\end{align*}
where 
the set of parameters $y_{l}$ satisfy Eq.(\ref{eq2}) with $C=0$, $B=2M-2-L$ and $\displaystyle A=({2G})^{-1}-\sum_{i=1}^{L}\varepsilon_{i}$. This case thus belong to case 2, as the previous example. We will see however, that the parameters $A$, $B$ and $C$ will generate very different behaviours for the roots of the generalised Heine-Stieltjes polynomials. The equation can be put into the form  
\begin{align*}
(z^{4}-\gamma z^{3}+\eta z^{2})Q''+\left(-\frac{\eta}{2G}+\frac{1}{2}\gamma \eta L+\left(2 \eta+\frac{\gamma}{2G}-\frac{(\gamma^{2}+2\eta) L}{2} -2 \eta M\right)\right. z     % \label{eq31}
\end{align*}
\[+\left.\left(-2 \gamma-\frac{1}{2G}+2 \gamma M\right) z^{2}+(2-2 M) z^{3}\right)Q'-(\beta_{2}z^{2}+\beta_{1}z+\beta_{0}  )Q=0.\]
We can show that the coefficients of the Van Vleck polynomial may be written as
\begin{align*}
\beta_{2}&=-\left((M-1)M\right),\quad \beta_{1}=-\left(\gamma M+\frac{M}{2G}-\gamma M^{2} \right) ,  \\
\beta_{0}&=-\left(\frac{E}{2G}+\frac{\gamma L}{4}-\eta M-\frac{\gamma M}{2G}+\frac{1}{2}(\gamma-\eta) L M+ \eta M^{2}   \right) .   
\end{align*}

For the numerical calculations we choose $\varepsilon_{2}=1$, $\varepsilon_{1}={1}/{2}$, $L=64$, $M=32$. The total number of states is given by $M$. Results are shown in Figs. 7-9.
These figures illustrate the changing behaviour of the roots near the point $g={2}/{3}$. The the ground-state roots form an arc in the regions near this point and collapse at the origin at the this point (similar behavior as that of the Moore-Read line of the Example 3). 
  
\begin{figure}
\centering
\subfloat[]{\includegraphics[width=5cm]{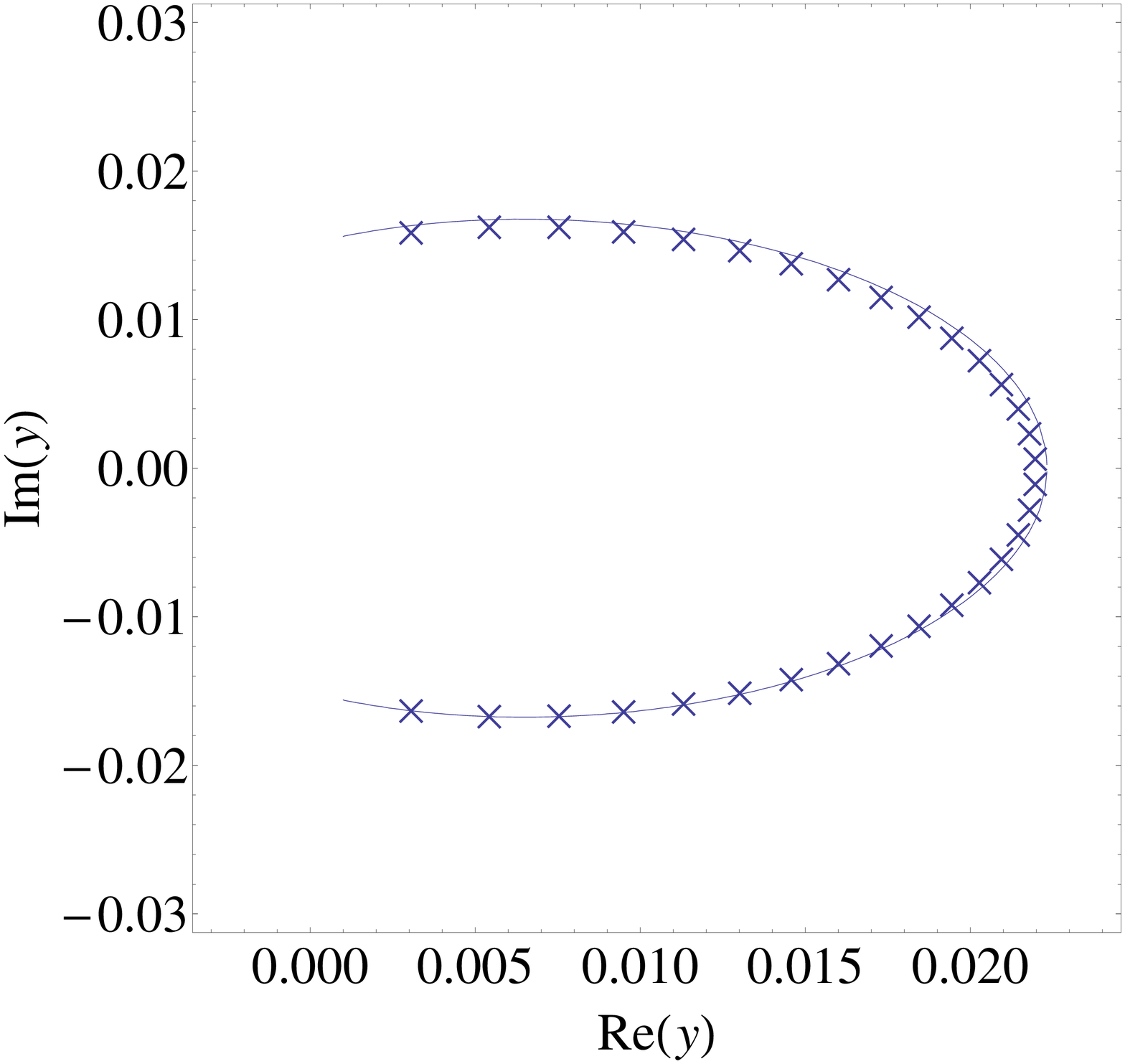}}
\subfloat[]{\includegraphics[width=5cm]{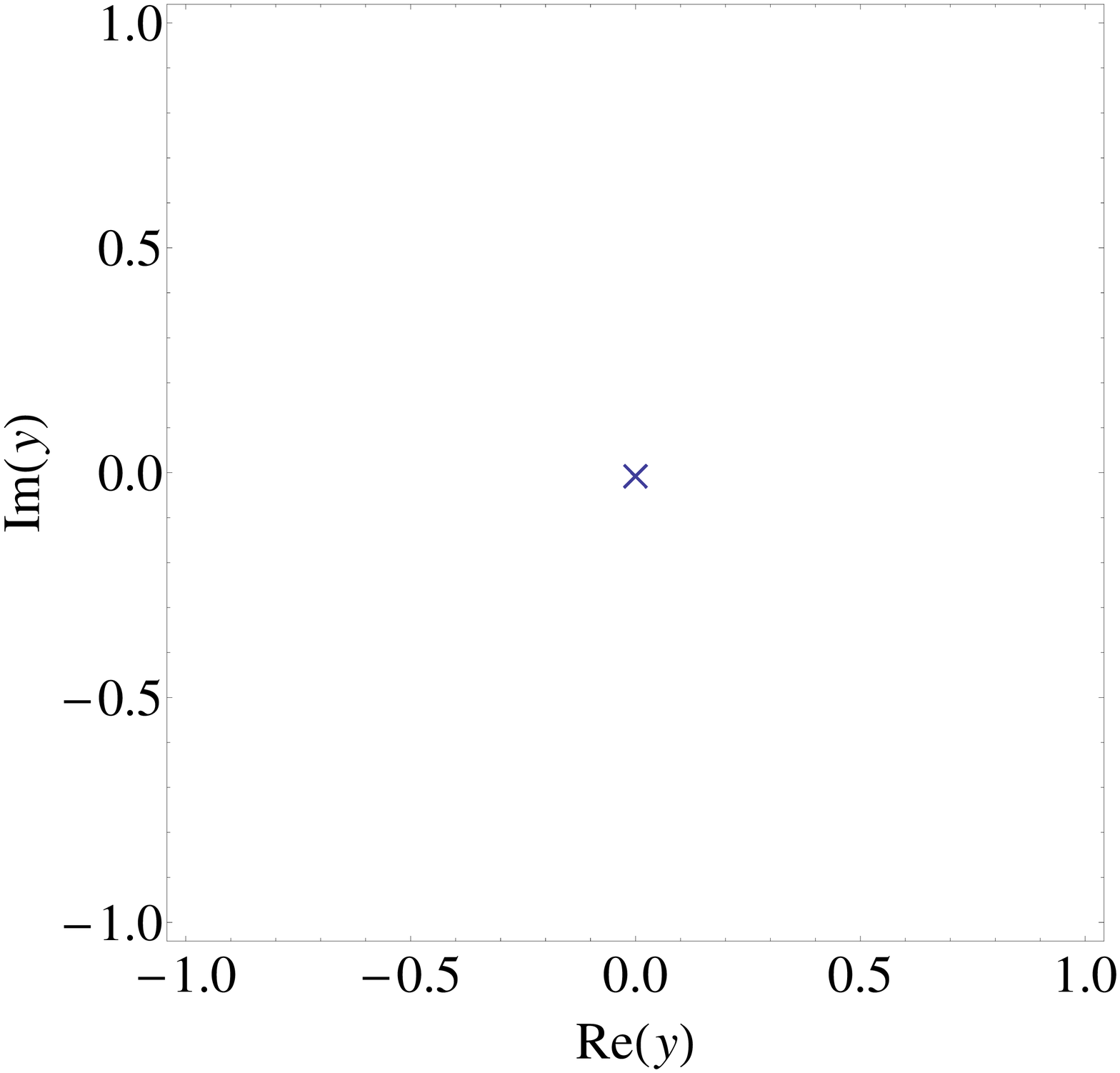}}
\subfloat[]{\includegraphics[width=5cm]{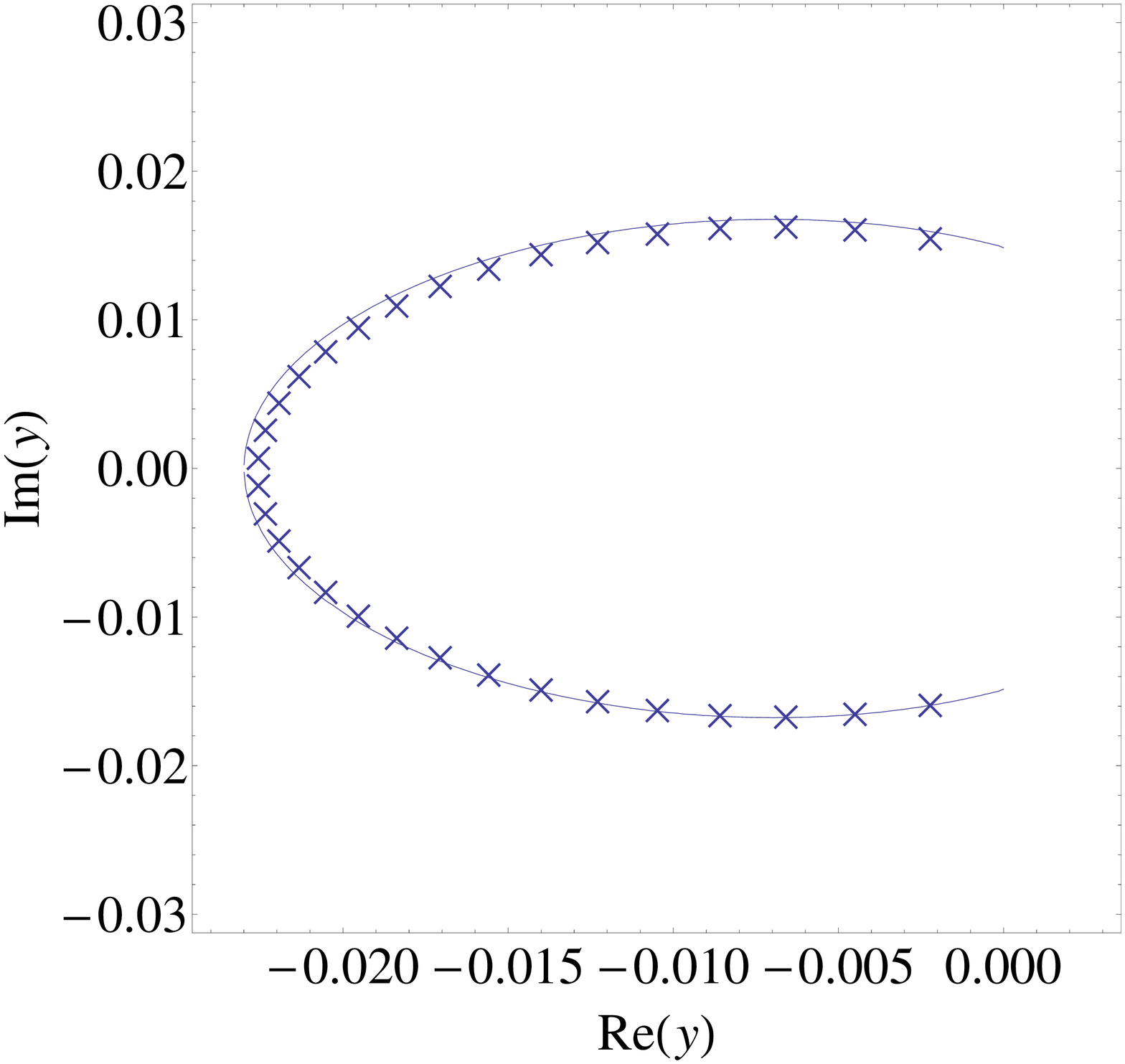}}
\caption{Roots for the ground state of the degenerate, two-level extended $d+id$-wave pairing model with $\varepsilon_{2}=1$,  $\varepsilon_{1}={1}/{2}$, $L=64$, and $M=32$: (a) $g={49}/{75}$ , (b) $g={2}/{3}$, and (c) $g={51}/{75}$. Also shown are the theoretical curves derived in Appendix B for the  large-$L$ limit. In case (b) the curve has contracted to a point at the origin.}
\label{fig1}
\end{figure} 

\begin{figure}
\centering
\subfloat[]{\includegraphics[width=6cm]{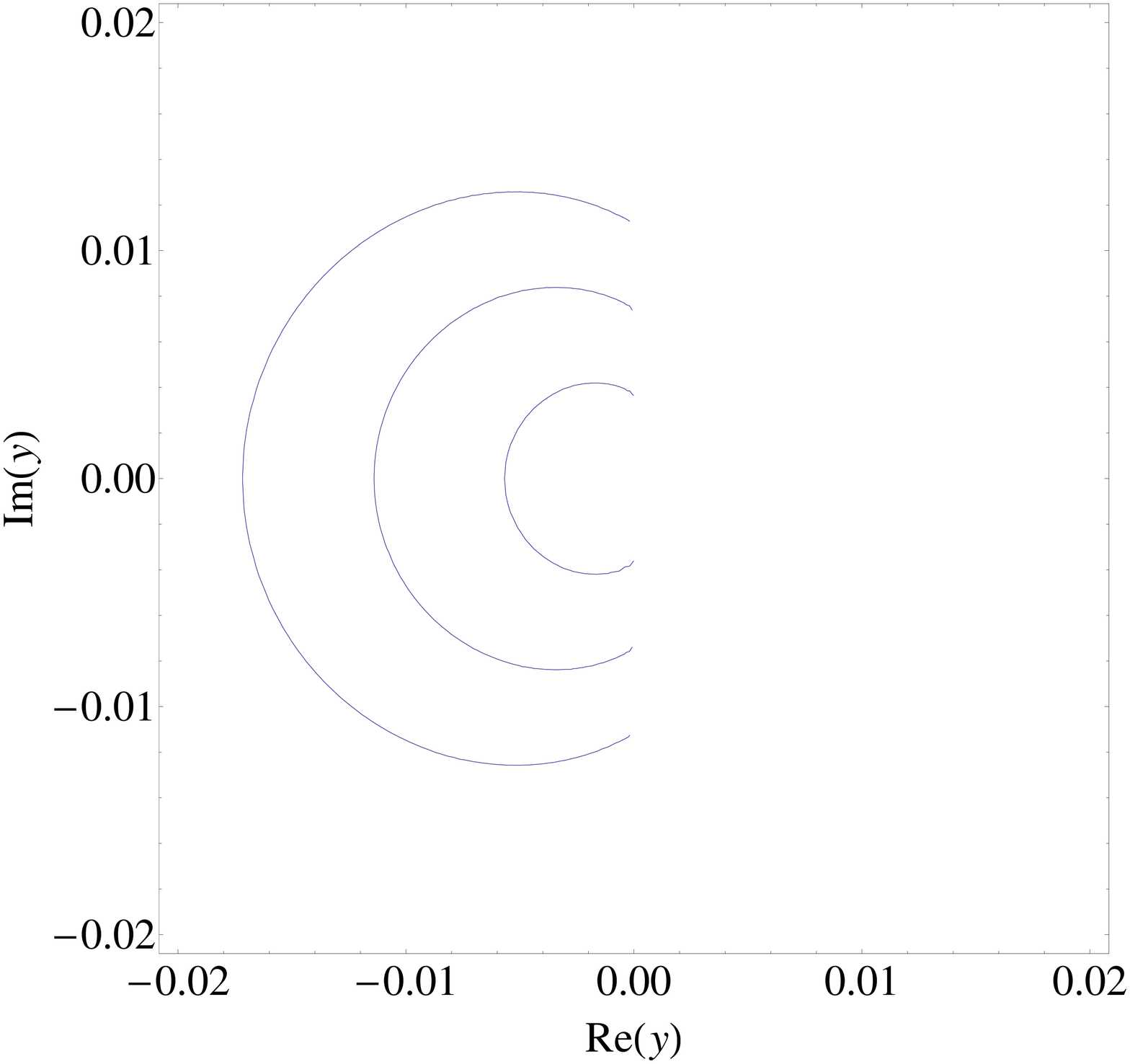}}
\subfloat[]{\includegraphics[width=6.1cm]{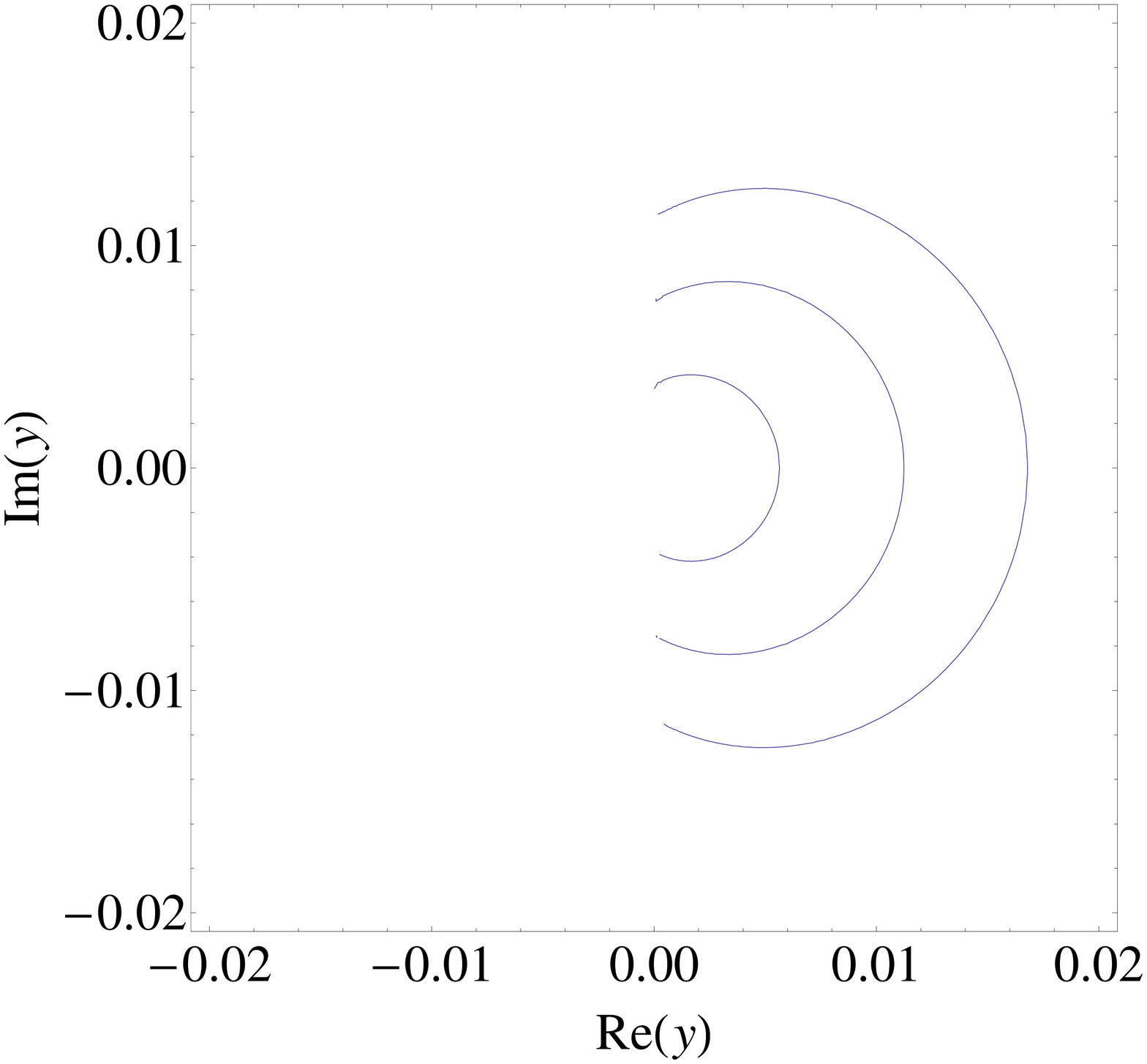}}
\caption{
Theoretical curves derived in Appendix B for the  large-$L$ limit of the degenerate, two-level extended $d+id$-wave pairing model with  $\varepsilon_{2}=1$, $\varepsilon_{1}={1}/{2}$, and 
$\displaystyle x = \lim_{L\rightarrow\infty} M/L=1/2$:  (a) From left to right 
$g = 203/300,\, g = 202/300, g = 201/300$, (b) From left to right $g = 199/300,\, g = 198/300,\,g = 197/300$. The limiting behaviour indicates that the curve contracts to a point at the origin when $g=2/3$.
}
\label{fig1}
\end{figure} 
  
\begin{figure}
\centering
\subfloat[]{\includegraphics[width=5cm]{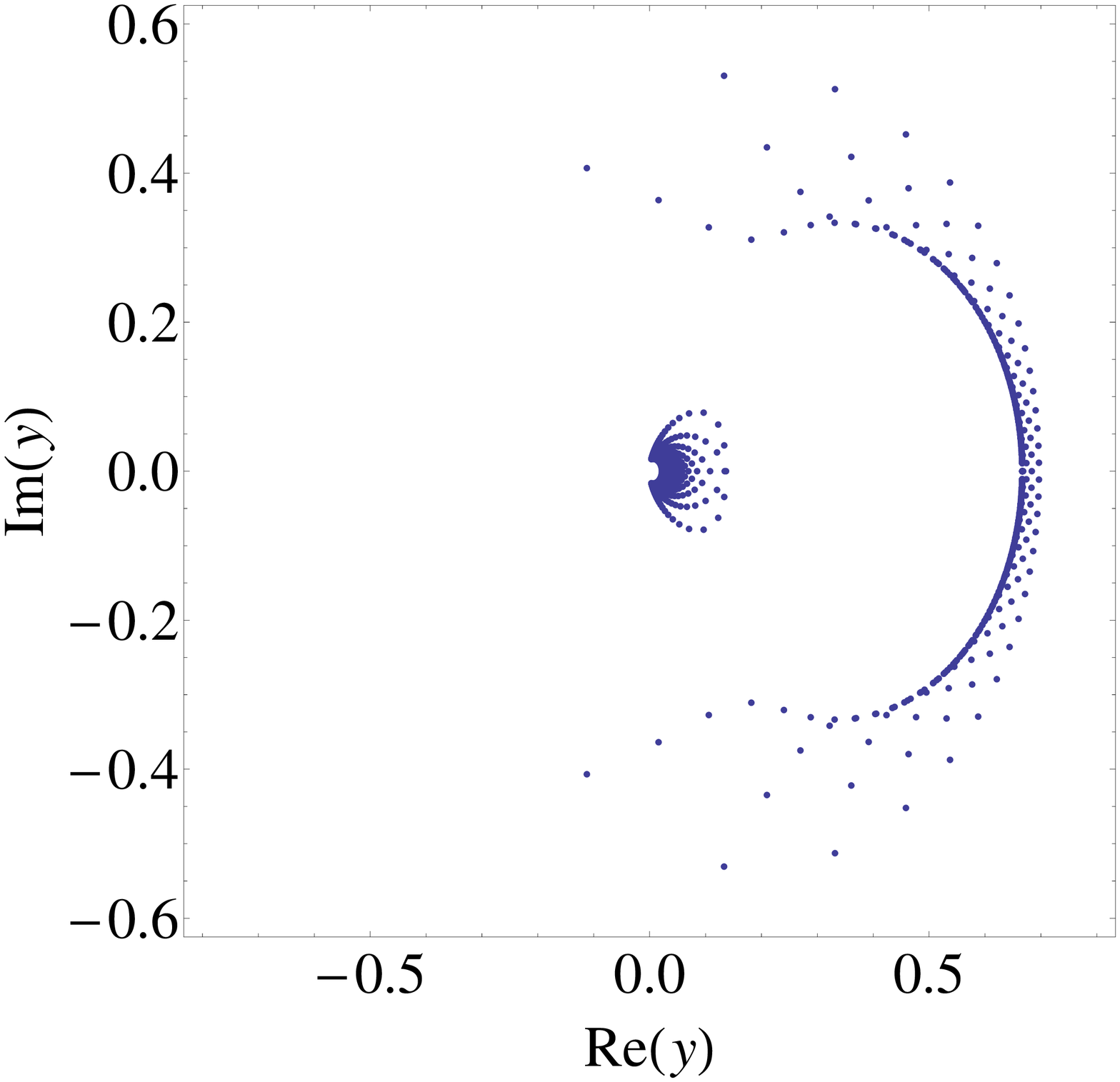}}
\subfloat[]{\includegraphics[width=5cm]{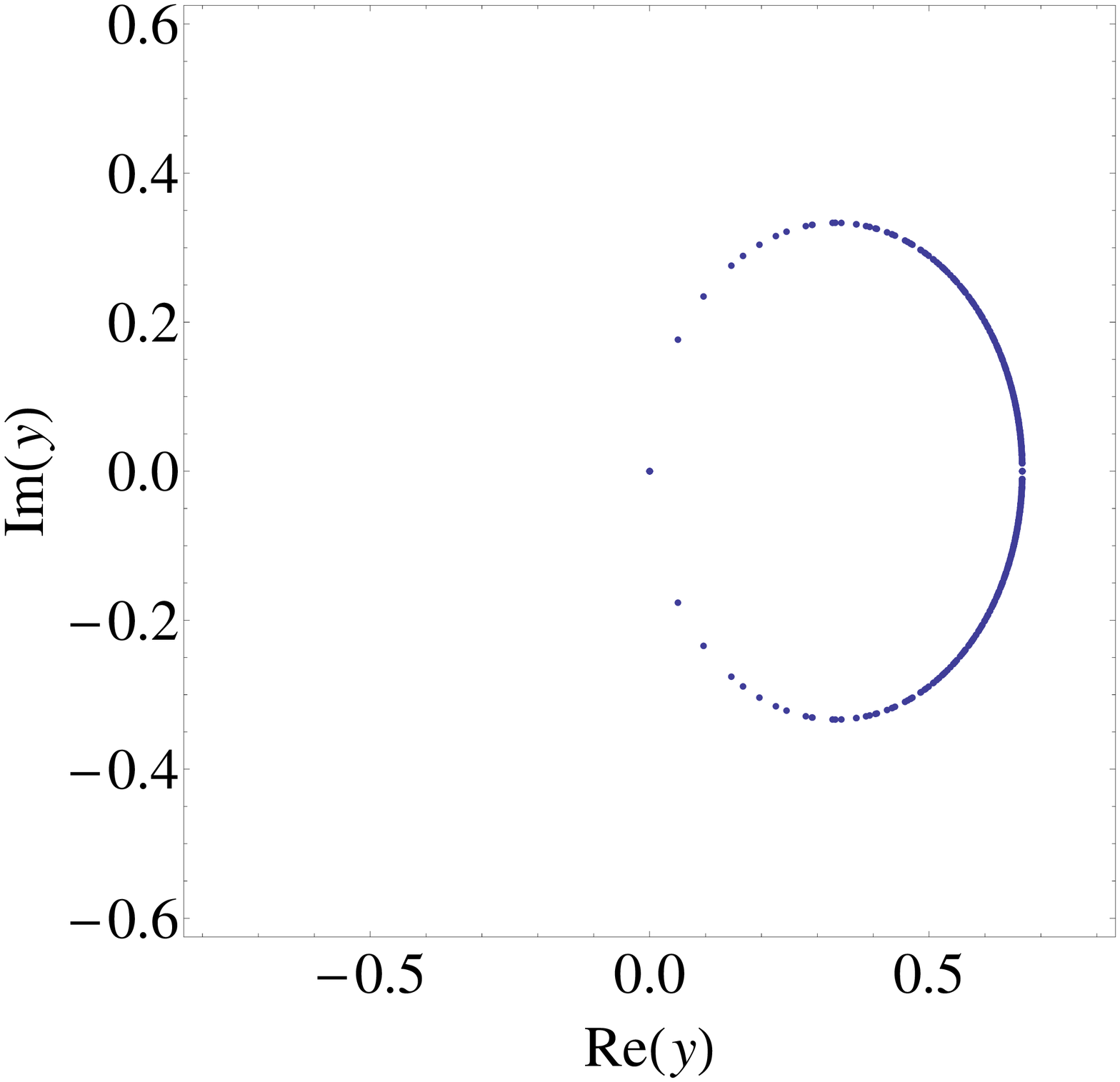}}
\subfloat[]{\includegraphics[width=5cm]{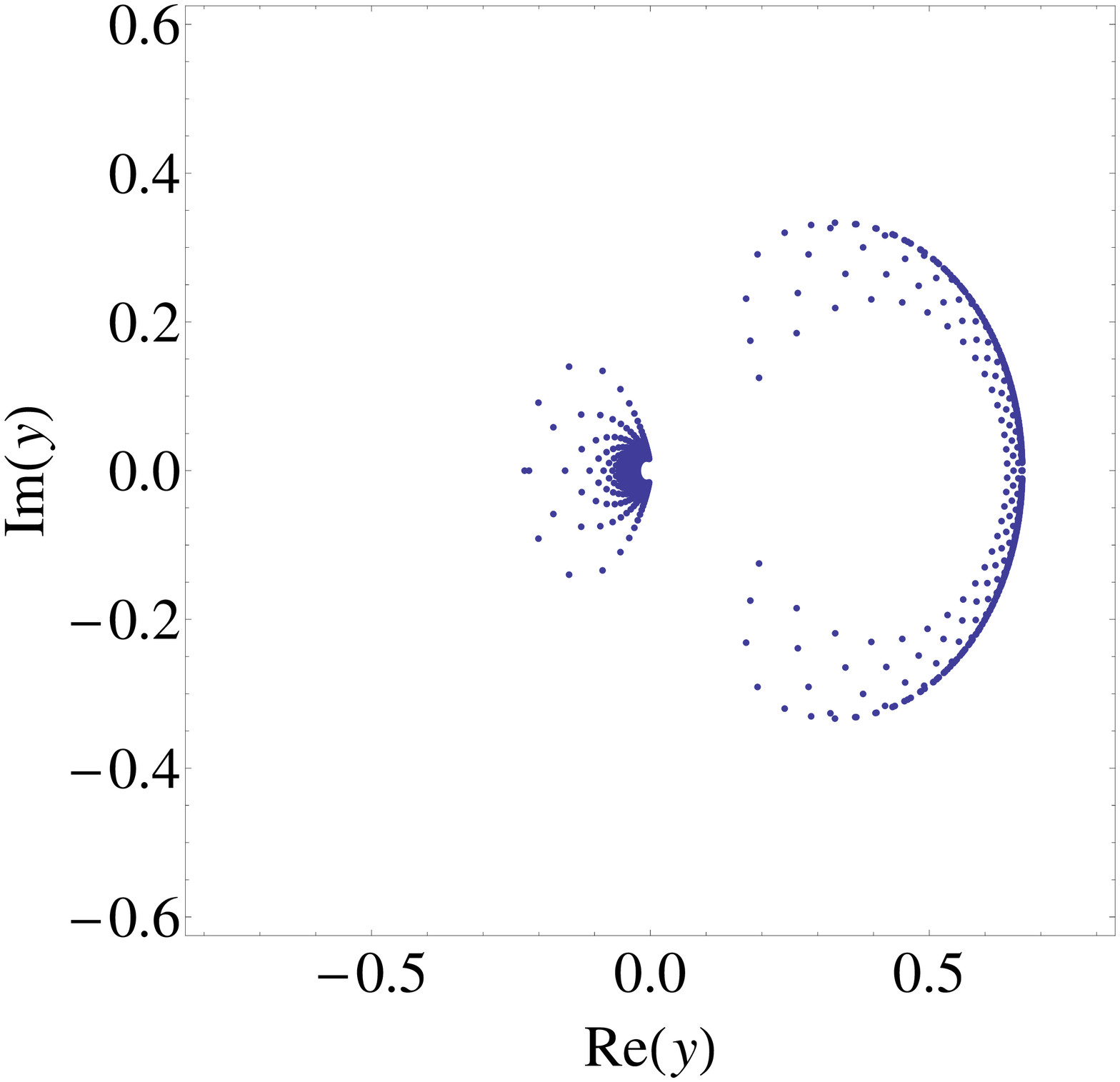}}
\caption{Roots for all unblocked symmetric states of the degenerate, two-level extended $d+id$-wave pairing model with $\varepsilon_{2}=1$, $\varepsilon_{1}={1}/{2}$, $L=64$, and $M=32$: (a) $g={49}/{75}$, (b) $g={2}/{3}$, and (c) $g={51}/{75}$. In each case a total of 33 sets of roots are displayed, where each set contains 32 roots.          }
\label{fig1}
\end{figure}

%Let us present the roots up to the fourth excited state 

%\begin{figure}
%\centering
%\subfloat[a]{\includegraphics[width=6cm]{Fig9a.eps}}
%\subfloat[b]{\includegraphics[width=6cm]{Fig9b.eps}}\\
%\subfloat[c]{\includegraphics[width=6cm]{Fig9c.eps}}
%\subfloat[d]{\includegraphics[width=6cm]{Fig9d.eps}}
%\caption{OMIT?????? The figure present the roots for the two-level integrable system for the ground state for %$\varepsilon_{2}=1$ and $\varepsilon_{1}=\frac{1}{2}$, $L=64$, $M=\frac{L}{2}$ and $g=\frac{2}{3}$ for
%(a) first excited state, (b) the second excited state, (c) the third excited state and (d) the fourth excited state}
%\label{fig1}
%\end{figure}

We finally identify four ground-state phase boundary lines which are associated with changes in the topology of the root distribution curve in the continuum limit. These cases are respectively shown by  Fig. 1(b) for the s-wave model, Fig. 3(b) and Fig. 3(d) for the $p+ip$-wave model, and Fig. 7(b) for the extended $d+id$-wave model. From the the numerical calculation using the BA/ODE correspondence, we find that the behaviour of these curves and the ground-state roots of the ground state can differ greatly. Remarkably, each of the cases show distinctive features which suggests that the four boundary lines should exhibit intrinsically different consequences for the properties of the models. The observations are summarised in Table 3. Another curious observation concerns the excited states. At the Read-Green line for the $p+ip$-pairing Hamiltonian as depicted in Fig. 4(d), and at the boundary for the extended $d+id$-pairing Hamiltonian  as depicted in Fig. 9(b), the roots for {\it all} excited states cluster near common curves.    
\begin{table}
\caption{Behaviour of the ground-state roots in a finite-sized system, and the distribution curves in the continuum limit, for ground-state phase boundary lines}
\footnotesize\rm
\begin{tabular*}{\textwidth}{@{}l*{15}{@{\extracolsep{0pt plus12pt}}l}}
\br
 Model & Figure   &   $\#$ roots at the origin &  Behavior of the curve          \\
\br
  $s$ &  1 (b)  &  Nil &   Closed/open transition          \\
  $p+ip$ (Moore-Read line) & 3 (b) & All &    Open/closed transition        \\
  $p+ip$ (Read-Green line) & 3 (d)  &  Nil &    Closed curve collapse      \\
 Extended  $d+id$ &  7 (b) &    All &   Open curve collapse/revival      \\
\mr
\end{tabular*}
\end{table}

\section{Conclusion}

We used the BA/ODE correspondence and the related generalised Heine-Stieltjes and Van Vleck polynomials to numerically obtain the roots of the Bethe Ansatz in the limit of two levels for four cases of integrable pairing models. We reproduced results obtain by other methods concerning the ground states, and we were able to study the behavior of the roots for the excited states as we change the coupling constant. 
The CPU time needed was consistent with what was mentioned in \cite{Pan1}. In this paper we used a monomial expansion and found that the method is robust and reliable in all cases with a moderate value of $M$ and a sufficient working precision. The application to general $N$-level models is straightforward, however, the CPU time needed to perform the calculations will increase. The most efficient method to obtain the roots of these polynomials in such cases needs further study. The existence of recursion operators as suggested in \cite{Pan2} is also an aspect to be investigated as it could lead to faster methods. This approach is not limited to integrable BCS models but can be applied to integrable systems that appear in other contexts, and to other classes of BA equations. 

An interesting aspect to investigate is the relation between the Schr\"'odinger form of the equations (\ref{eq7}) for these four cases, and quasi-exactly solvable systems (QES) \cite{Gon1}. The classification of QES systems was performed for certain classes of equations of the form given by (\ref{eq7}) with polynomials of order 2,3 and 4. The equations in this paper seem to be beyond this classification, and also other systems obtained more recently \cite{Gon1,Aoy,Gon2}. Supersymmetric quantum mechanics appears to be useful method to obtain such systems \cite{Aoy,Gon2}. 

\ack This work was supported by the Australian Research Council through Discovery Project DP110101414.

\renewcommand{\theequation}{A.\arabic{equation}}    
  % redefine the command that creates the equation no.    
  \setcounter{equation}{0}  % reset counter     
\section*{Appendix A. Explicit generalised Heine-Steiltjes and Van Vleck polynomials}
The following cases correspond to the four Examples of Sect. 3. It is seen that the coefficients can differ across many orders of magnitude. The numerical precision used in the computation was significantly higher than the number of digits shown below.  
\newline
\newline
\textbf{Example 1: ground state at $g=1$}
\begin{align*}
\alpha_{0}&= 4.3,                & \alpha_{1}&=2.1 \times 10^{2}, &
\alpha_{2}&=4.9 \times 10^{3},   & \alpha_{3}&=7.3 \times 10^{4}, \\
\alpha_{4}&=8.1 \times 10^{6},   & \alpha_{5}&=7.1 \times 10^{6}, &
\alpha_{6}&=5.1 \times 10^{7},   & \alpha_{7}&=3.1 \times 10^{8}, \\
\alpha_{8}&=1.6 \times 10^{9},   & \alpha_{9}&=7.1 \times 10^{9},  &
\alpha_{10}&=2.8 \times 10^{10}, &\alpha_{11}&=9.8 \times 10^{10}, \\
\alpha_{12}&=3.1 \times 10^{11}, &\alpha_{13}&=8.6 \times 10^{11}, & 
\alpha_{14}&=2.2 \times 10^{12}, &\alpha_{15}&=5.1 \times 10^{12}, \\
\alpha_{16}&=1.1 \times 10^{13}, &\alpha_{17}&=2.1 \times 10^{13}, & 
\alpha_{18}&=3.7 \times 10^{13}, &\alpha_{19}&=6.0 \times 10^{13}, \\
\alpha_{20}&=9.1 \times 10^{13}, &\alpha_{21}&=1.3 \times 10^{14}, &
\alpha_{22}&=1.6 \times 10^{14}, &\alpha_{23}&=1.9 \times 10^{14}, \\ 
\alpha_{24}&=2.1 \times 10^{14}, &\alpha_{25}&=2.1 \times 10^{14}, & 
\alpha_{26}&=2.0 \times 10^{14}, &\alpha_{27}&=1.7 \times 10^{14}, \\
\alpha_{28}&=1.4 \times 10^{14}, &\alpha_{29}&=1.0 \times 10^{14}, &
\alpha_{30}&=7.0 \times 10^{13}, &\alpha_{31}&=4.4 \times 10^{13}, \\
\alpha_{32}&=2.6  \times10^{13}, &\alpha_{33}&=1.4  \times 10^{13}, &
\alpha_{34}&=6.7 \times 10^{12}, &\alpha_{35}&=3.0 \times 10^{12}, \\
\alpha_{36}&=1.2  \times10^{12}, &\alpha_{37}&=4.5 \times 10^{11}, & 
\alpha_{38}&=1.5 \times 10^{11}, &\alpha_{39}&=4.6 \times 10^{10}, \\
\alpha_{40}&=1.2  \times10^{10}, &\alpha_{41}&=2.9 \times 10^{9}, &
\alpha_{42}&=6.2 \times 10^{8},  &\alpha_{43}&=1.1 \times 10^{8}, \\
\alpha_{44}& =1.8  \times10^{7}, &\alpha_{45}&=2.3 \times 10^{6}, & 
\alpha_{46}&=2.5 \times 10^{5}, & \alpha_{47}&=2.1 \times 10^{4}, \\
& &  \alpha_{48}&=1.3 \times 10^{3}, & \alpha_{49}&=5.1 \times 10^{4}, 
\end{align*}
\begin{align*}
\beta_{0}&=2.5 \times 10^{3}, &  \beta_{1}&=-5000.
\end{align*}
\newline
The coefficient $\beta_{1}$ is an exact value which is independent of the state.
\newline
\newline
\textbf{Example 2: ground state at $g=3/2$.}
\begin{align*}
\alpha_{0}&= 0, &\alpha_{1}&=0, &\alpha_{2}&=0, & \alpha_{3}&=0, \\ 
\alpha_{4}&=0,  & \alpha_{5}&=0, & 
\alpha_{6}&=0, &\alpha_{7}&=0, \\
\alpha_{8}&=0, &\alpha_{9}&=0, & \alpha_{10}&=0, &  \alpha_{11}&=0, \\
\alpha_{12}&=-1.7 \times 10^{-46}, & \alpha_{13}&=-6.5 \times 10^{-45}, & \alpha_{14}&=-2.2 \times 10^{-43}, &
 \alpha_{15}&=7.3 \times 10^{-42}, \\
  \alpha_{16}&= -2.3 \times 10^{-40}, &\alpha_{17}&=6.9 \times 10^{-39},& 
\alpha_{18}&=-2.0 \times 10^{-37}, & \alpha_{19}&=5.8 \times 10^{-36}, \\
\alpha_{20}&=-1.6 \times 10^{-34}, & \alpha_{21}&=4.6 \times 10^{-33}, &\alpha_{22}&=-1.3 \times 10^{-31}, & 
\alpha_{23}&=3.5 \times 10^{-30}, \\
\alpha_{24}&=-1.0 \times 10^{-28},  &\alpha_{25}&=2.9 \times 10^{-27}, & \alpha_{26}&=-8.9 \times 10^{-26}, & 
\alpha_{27}&=2.9 \times 10^{-24}, \\
\alpha_{28}&=-1.0 \times 10^{-22}, & \alpha_{29}&=4.1 \times 10^{-21}, & \alpha_{30}&=-1.9 \times 10^{-19}, & 
\alpha_{31}&=1.2 \times 10^{-17}, \\
\alpha_{32}&=-1.2 \times 10^{-15}, & \alpha_{33}&=4.3 \times 10^{-13}, & \alpha_{34}&=7.1 \times 10^{-11},  &
\alpha_{35}&=4.3 \times 10^{-9}, \\
\alpha_{36}&=1.4 \times 10^{-7}, &\alpha_{37}&=3.2 \times 10^{-6}, & \alpha_{38}&=5.0 \times 10^{-5}, & 
\alpha_{39}&=5.7 \times 10^{-4}, \\
\alpha_{40}&=5.0 \times 10^{-3}, & \alpha_{41}&=3.3 \times 10^{-2}, & \alpha_{42}&=1.7 \times 10^{-1}, & 
\alpha_{43}&=6.6 \times 10^{-1}, \\
\alpha_{44}&=2.0,  &\alpha_{45}&=4.7, &  \alpha_{46}&=8.1, & \alpha_{47}&=1.0 \times 10^{1}, \\
& & \alpha_{48}&=8.3,  &\alpha_{49}&=4.3, 
\end{align*}
\begin{align*}
\beta_{0}&=6.8 \times 10^{4}, &  \beta_{1}&=-{27500}/{3}.
\end{align*}
The coefficient $\beta_{1}$ is an exact value which is independent of the state.
\newline
\newline
\textbf{Example 3: ground state at $g=1$.}
\begin{align*}
\alpha_{0}&= 2.4 \times 10^{-11}, & \alpha_{1}&= 2.3 \times 10^{-9}, &\alpha_{2}&= 1.1 \times 10^{-7}, &\alpha_{3}&= 2.8 \times 10^{-6}, \\
\alpha_{4}&= 0.0, &  \alpha_{5}&=6.0 \times 10^{-4}, &\alpha_{6}&=6.1 \times 10^{-3}, &\alpha_{7}&= 0.043, \\
\alpha_{8}&= 0.23, & \alpha_{9}&=9.4 \times 10^{-1}, &\alpha_{10}&= 2.9, &\alpha_{11}&= 6.6, \\
\alpha_{12}&= 11,  &\alpha_{13}&=13,  &\alpha_{14}&= 10, & \alpha_{15}&= 4.8, 
\end{align*}
\begin{align*}
\beta_{0}&=-2.0 \times 10^{2}, &\beta_{1}&=8.6 \times 10^{2}, & \beta_{2}&=-768.
\end{align*}
The coefficient $\beta_{2}$ is an exact value which is independent of the state. 
\newline
\newline
\textbf{Example 4: ground state at $g={102}/{150}$.}
\begin{align*}
\alpha_{0}&=6.1 \times 10^{-55}, & \alpha_{1}&=6.5 \times 10^{-52}, & \alpha_{2}&=3.5 \times 10^{-49}, & \alpha_{3}&=1.3 \times 10^{-46}, \\
\alpha_{4}&=3.3 \times 10^{-44}, & \alpha_{5}&=7.1 \times 10^{-42}, & \alpha_{6}&=1.3 \times 10^{-39}, & \alpha_{7}&=1.9 \times 10^{-37}, \\
\alpha_{8}&=2.4 \times 10^{-35}, & \alpha_{9}&=2.7 \times 10^{-33}, & \alpha_{10}&=2.7 \times 10^{-31}, & 
\alpha_{11}&=2.4 \times 10^{-29}, \\
\alpha_{12}&=1.9 \times 10^{-27}, & \alpha_{13}&=1.4 \times 10^{-25}, & \alpha_{14}&=9.0 \times 10^{-24}, &
\alpha_{15}&=5.3 \times 10^{-22}, \\
\alpha_{16}&=2.8 \times 10^{-20}, & \alpha_{17}&=1.4 \times 10^{-18}, & \alpha_{18}&=5.9 \times 10^{-17}, & 
\alpha_{19}&=2.3 \times 10^{-15}, \\
\alpha_{20}&=8.2 \times 10^{-14}, & \alpha_{21}&=2.6 \times 10^{-12}, & \alpha_{22}&=7.3 \times 10^{-11}, & 
\alpha_{23}&=1.8 \times 10^{-9}, \\
\alpha_{24}&=4.0 \times 10^{-8},  & \alpha_{25}&=7.7 \times 10^{-7} , & \alpha_{26}&=0.0, & \alpha_{27}&=1.0 \times 10^{-4}, \\  \alpha_{28}&=1.9 \times 10^{-3}, & \alpha_{29}&=1.7 \times 10^{-2}, & \alpha_{30}&=1.1 \times 10^{-1}, & 
\alpha_{31}&=4.6 \times 10^{-1},  
\end{align*}
\begin{align*}
\beta_{0}&=5.1 \times 10^{2}, & \beta_{1}&=-{304}/{17}, & \beta_{2}=-1.0 \times 10^{3}.
\end{align*}
The coefficients $\beta_{1}$ and $\beta_2$ are both exact values which are independent of the state.

\renewcommand{\theequation}{B.\arabic{equation}}    
  % redefine the command that creates the equation no.    
  \setcounter{equation}{0}  % reset counter     
\section*{Appendix B. Solution of the Bethe Ansatz equations in the continuum limit}
If in the limit $L\rightarrow \infty$ the roots of the Bethe Ansatz equations (\ref{eq2}) are densely distributed on a curve $\Gamma$ in the complex plane, we can look to find a solution via integral equation methods. We refer to this as the continuum limit. Specifically we obtain the singular integral equation 
\begin{equation}
\frac{A}{y^{2}}+\frac{B}{y}+C + \int_{\Omega}d\varepsilon  \frac{\rho(\varepsilon)}{y-\varepsilon}+P\int_{\Gamma}|dy'| \frac{2r(y')}{y'-y}=0  
\label{int}
\end{equation}
where $\Omega$ denotes the interval of the real line where the energy levels $\varepsilon$ lie, distributed according to a density $\rho(\varepsilon)$ such that
\begin{align*}
\int_{\Omega}d\varepsilon\, \rho(\varepsilon)=1.
\end{align*}
These integral equations may be solved using complex analysis techniques (cf. \cite{Gau,Roma,Ami2,Dun2,Romb,ml12}). Since the energy must be real, we assume that the arc $\Gamma$ is invariant under reflections about the real axis. The solution for  $\Gamma$ having end points  $a=\epsilon - i \delta$ and $b=\epsilon + i \delta$ is of the form
\begin{align*}
r(y)|dy|&=\frac{1}{2\pi i}(h_{+}(y)-h_{-}(y))dy,\\ 
h(y)&= R(y)\left( \int_{\Omega} d\varepsilon \frac{\phi(\varepsilon)}{\varepsilon-y}+u_{0}+\frac{u_{1}}{y}+\frac{u_{2}}{y^{2}}\right), \\ R(y)&=\sqrt{(y-a)(y-b)}=\sqrt{(y-\epsilon)^{2}+\delta^{2}}.
\end{align*}
The Cauchy principal value in Eq. (\ref{int}) can be written as:
\begin{align*}
P\int_{\Gamma}2|dy'|\frac{r(y')}{y'-y}=\oint_{C_{\Gamma}}\frac{dy'}{2\pi i}\frac{h(y')}{y'-y},
\end{align*} 
where $C_{\Gamma}$ is a closed curve which contains $\Gamma$ in the interior.
Similar equations apply for the case where $\Gamma$ lies on the real axis with support on the interval $[a,b]$.

The solution can thus be obtained on the form of contraints that involve integrals of the density $\rho(\varepsilon)$. For all four cases
relations of the form 
\begin{equation}
x={\mathcal F}_{1}(a,b),\quad g={\mathcal F}_{2}(a,b)
\label{xg}
\end{equation}
can be obtained. For the two-level models we use $\displaystyle \rho(\varepsilon)=\delta(\varepsilon-\epsilon_{1})/2 +\delta(\varepsilon-\epsilon_{2})/2$, and for a given
$g$ and $x$ we may numerically determine the end points of $\Gamma$. From these we use the equation
\begin{align*}
{\rm Re}\left[ \int_{a}^{\xi}d\xi'\,h(\xi')\right]=0
\end{align*}
to obtain the arc.  

Alternatively, the arc may form a closed curved in the complex plane. In this case the solution of the singular integral equation takes the form
\begin{align*}
s(y)= \frac{1}{i\pi}\left( \int_{\Omega} d\varepsilon \frac{\rho(\varepsilon)}{\varepsilon-y}+v_{0}+\frac{v_{1}}{y}+\frac{v_{2}}{y^{2}}
\right)
\end{align*}
and the curve is given by
\begin{align*}
{\rm Im}\left[ \int_{a}^{\xi}d\xi'\,s(\xi')\right]=0.
\end{align*}
Here, the introduction of a integration constant is necessary. It can be obtained by imposing the constraint that the value $w$ such $s(w)=0$ lies on the curve.
In the case that $\Gamma=\Gamma_1 \cup \Gamma_2$, where $\Gamma_1$ is a closed curve that touches a point $b$ of $\Gamma_{2}=(a,b) \nsubseteq \Omega$ we have
\begin{align*}
s(y)=\frac{\sqrt{(y-a)(y-b)}}{i\pi}\left( \int_{\Omega} d\varepsilon \frac{\phi(\varepsilon)}{\varepsilon-y}+v_{0}+\frac{v_{1}}{y}+\frac{v_{2}}{y^{2}}
\right).
\end{align*}

Below we provide details for when the arc $\Gamma$ is associated to the ground-state roots in each of the four examples. For each case it is found that $\phi(\varepsilon)=\rho(\varepsilon)/|R(\varepsilon)|$.
\newline
\newline
\textbf{Example 1}
\newline
\newline
The corresponding equation for Eq. (\ref{xg}) are
\begin{align*}
x&=\frac{1}{2}\int_{\Omega}d\varepsilon\,\rho(\varepsilon)\left(1-\frac{\varepsilon-\epsilon}{R(\varepsilon)}\right) , \\
\frac{1}{g}&=\int_{\Omega}d\varepsilon\,\frac{\rho(\varepsilon)}{R(\varepsilon)}
\end{align*}
with $u_{0}=u_{1}=u_{2}=0$. In the case of closed curve we have $v_{2}=v_{1}=0$ and $v_{0}={1}/{(2g)}$. In this case explicit expressions for the curve can be obtained for the open arc \cite{Roma}
\begin{align*}
x^{2}+y^{2}+g^{2}=\frac{2xg}{\tanh({2x}{g}^{-1})},\quad  g \geq \varepsilon_{1},
\end{align*}
and for the closed arc
\begin{align*}
x^{2}+y^{2}+\varepsilon_{1}^{2}=\frac{2x\varepsilon_{1}}{\tanh({2(x-\xi_{0})}{g^{-1}})},  \quad g  < \varepsilon_{1}
\end{align*}
with
\begin{align*}
\xi_{0}=\chi_{0}-\frac{g}{2}\ln\left(\frac{\varepsilon_{1}+\chi_{0}}{\varepsilon_{1}-\chi_{0}}\right),\quad \chi_{0}=-\varepsilon_{1}\sqrt{1-\frac{g}{\varepsilon_{1}}}.
\end{align*}
\newline
\newline
\textbf{Example 2}
\newline
\newline
The corresonding equations for (\ref{xg}) are
\begin{align*}
2x&=1-\frac{1}{g}+\sqrt{\epsilon^{2}+\delta^{2}}\int_{\Omega}d\varepsilon\,\frac{\rho(\varepsilon)}{R(\varepsilon)} , \\
\frac{1}{g}&=\int_{\Omega}d\varepsilon\, \frac{\varepsilon \rho(\varepsilon)}{R(\varepsilon)} ,
\end{align*}
with $u_{0}=u_{2}=0$  and $u_{1}=(2x-1+g^{-1})(\epsilon^{2}+\delta^{2})^{{3}/{2}}/2$. In the case of a closed curve we have $v_{0}=v_{2}=0$ and 
$v_{1}=2x-1+g^{-1}$. In the case of a closed curve $\Gamma_1$ that touches a point $b$ of $\Gamma_{2}=(a,b) \nsubseteq \Omega$ we have $v_{0}=v_{2}=0$ and 
$v_{1}=(2x-1+g^{-1})/ab$.
\newline
\newline
\textbf{Example 3}
\newline
\newline
In this case the arc associated with the ground-state roots is always an interval $[a,b]$ on the real axis.
The corresponding equations for (\ref{xg})  are
\begin{align*}
q+\frac{f^{2}(a+b)}{2\sqrt{ab}}&=-\sqrt{ab}\int_{\Omega}d\varepsilon\, \frac{\rho(\varepsilon)}{2R(\varepsilon)}, \\
\frac{1}{g}&=\frac{f^{2}}{\sqrt{ab}}+\int_{\Omega}d\varepsilon\, \frac{\varepsilon \rho(\varepsilon)}{2R(\varepsilon)}
\end{align*}
with
\begin{align*}
q&=2x-1+\frac{1}{g}, \\
f&=\frac{F}{\sqrt{L}}, \\
u_{0}&=0, \\
u_{1}&=-\frac{1}{2\sqrt{ab}}\left(q+\frac{f^{2}(a+b)}{4ab}\right), \\
u_{2}&=-\frac{f^{2}}{2\sqrt{ab}}.
\end{align*}
\newline
\newline
\textbf{Example 4}
\newline
\newline
In this case the equations given by (\ref{xg}) take the form
\begin{align*}
x&=\frac{1}{2}+\frac{\sqrt{\epsilon^{2}+\delta^{2}}}{2}\int_{0}^{\omega}d\varepsilon\, \phi(\varepsilon) -\frac{\epsilon}{2\sqrt{\epsilon^{2}+\delta^{2}}}\int_{0}^{\omega}d\varepsilon\, \phi(\varepsilon) \varepsilon , \\
\frac{1}{g}&=2\int_{0}^{\omega}d\varepsilon\,\varepsilon \rho(\varepsilon)   + 2\sqrt{\epsilon^{2}+\delta^{2}} \int_{0}^{\omega} d\varepsilon\,\varepsilon \phi(\varepsilon)  
\end{align*}
with
\begin{align*}
c&=2x-1, \\
f&=\frac{1}{2g}-\int_{\Omega}d\varepsilon\,\varepsilon \rho(\varepsilon), \\
u_{0}&=0, \\
u_{1}&=\frac{c}{\sqrt{\epsilon^{2}+\delta^{2}}}+\frac{f\epsilon}{(\epsilon^{2}+\delta^{2})^{{3}/{2}}}, \\
u_{2}&=\frac{f}{\sqrt{\epsilon^{2}+\delta^{2}}}.
\end{align*}.

\end{document}